\documentclass[referee]{aa} % for a referee version
\newcommand{\kei}{K$_s$-} 
\newcommand{\gei}{$J$-}
\newcommand{\KEI}{K$_s$} 
\newcommand{\GEI}{$J$}
\newcommand{\xbong}{{\tt {XBONG}}}
\newcommand{\xbongs}{{\tt {XBONGs}}}
\def\xmm{{XMM-{\it Newton}}}
\def\chandra{{\it Chandra}}
\newcommand{\cgs}{ ${\rm erg~cm}^{-2}~{\rm s}^{-1}$} 
\newcommand{\lum}{\rm erg~s$^{-1}$}
\def\gtrsim{\mathrel{\hbox{\rlap{\hbox{\lower4pt\hbox{$\sim$}}}\hbox{$>$}}}}

\def\lum{{\rm {erg~s$^{-1}$}}}
\usepackage{graphicx}
\usepackage{txfonts}
\usepackage{natbib}
\bibpunct{(}{)}{;}{a}{}{,}
\begin{document}

   \title{The HELLAS2XMM survey: XI. Unveiling the nature of X-ray Bright Optically Normal Galaxies 
\footnote{Based on observations made at the 
   European Southern Observatory, Paranal, Chile (ESO Programme ID 69.A-0554).}}

   \author{ F. Civano\inst{1,}\inst{2}, M. Mignoli\inst{2}, A. Comastri\inst{2}, 
   C. Vignali\inst{1,}\inst{2}, 
   F. Fiore\inst{3}, L. Pozzetti\inst{2}, M. Brusa\inst{4}, F. La Franca\inst{5}, G. Matt\inst{5}, 
    S. Puccetti\inst{3,}\inst{6},F. Cocchia\inst{7,}\inst{3}}
   %\thanks{E-mail: francesca.civano@bo.astro.it}, 
 %  M. Mignoli\inst{2}, A. Comastri\inst{2}, C. Vignali\inst{1}, F. Fiore\inst{3}, F. Cocchia\inst{4}, et al.}

   \institute{Dipartimento di Astronomia, Universit\`a di Bologna,
    via Ranzani 1, 40127 Bologna, Italy 
   \and
      INAF - Osservatorio Astronomico di Bologna,
    via Ranzani 1, 40127 Bologna, Italy
   \and
    INAF - Osservatorio Astronomico di Roma, via Frascati 33, I-00040 Monteporzio-Catone (RM) , Italy
    \and
    Max Planck Institut fuer Extraterrestrische Physik (MPE), Giessenbachstr. 1, 85748 Garching, Germany
    \and
    Dipartimento di Fisica, Universit\`a Roma Tre, via della Vasca Navale 84, 00146 Roma, Italy
    \and
    ASI Science Data Center, ASDC c/o ESRIN, via G. Galilei, 00044 Frascati, Italy
    \and
    INAF - Osservatorio Astronomico di Brera, via Brera 28, I-20121 Milano, Italy 
    }

   \date{Accepted October 3, 2007}
\abstract{}{X-ray Bright Optically Normal Galaxies (\xbongs) constitute a small but not negligible fraction
of hard X-ray selected sources in recent \chandra\ and \xmm\
surveys. Even though several possibilities were proposed to explain
why a relatively luminous hard X-ray source does not leave any
significant signature of its presence in terms of optical emission
lines, the nature of \xbongs\ is still subject of debate. We aim to a
better understanding of their nature by means of a
multiwavelength and morphological analysis of a small sample of these
sources. } {Good-quality photometric near-infrared data (ISAAC/VLT) of
four low-redshift ($z=0.1-0.3$) \xbongs, selected from the {\rm HELLAS2XMM} survey,
have been used to search for the presence of the putative nucleus,
applying the surface-brightness decomposition technique through 
the least-squares fitting program GALFIT. } 
{The surface brightness decomposition allows us to
reveal a nuclear point-like source, likely to be responsible of the
X-ray emission, in two out of the four sources.  The results indicate that
moderate amounts of gas and dust, covering a large solid angle
(possibly 4$\pi$) at the nuclear source, combined with the low nuclear
activity, may explain the lack of optical emission lines. The third \xbong\ is associated with an
X-ray extended source and no nuclear excess is detected in the
near infrared at the limits of our observations. The last source is
associated to a close (d$\leq$ 1 arcsec) double system and the fitting
procedure cannot achieve a firm conclusion.}{}

    \keywords{galaxies: active - X-rays:galaxies}
   
	\authorrunning{F. Civano et al.}
	\titlerunning{Unveiling the nature of XBONGs}
	   \maketitle
%
%________________________________________________________________

\section{Introduction}

Thanks to the \chandra\ and \xmm\ surveys, the hard
X-ray sky is now probed down to a flux limit where the bulk of the
X-ray background is almost completely resolved into discrete sources 
(Hasinger et al. 2001; Alexander et al. 2003; Bauer et al. 2004; Worsley et al. 2004, 2005).  
Extensive programs of multiwavelength follow-up observations showed that
the large majority of hard X-ray selected sources are identified with
Active Galactic Nuclei (AGN) spanning a broad range of redshifts 
and luminosities. At variance with optically selected quasars,    
X-ray selected AGN are characterized by a much larger spread in their 
optical properties, especially for what concerns the intensity of the emission lines.
Indeed, a sizable fraction of relatively luminous X-ray sources 
hosting an active nucleus 
would not have been easily recognized as such 
on the basis of optical observations either because associated 
with very faint ($R >$ 24) counterparts (e.g., Fiore et al. 2003; 
Mignoli et al. 2004; Civano et al. 2005) 
or due to the lack of AGN emission lines in their optical spectra.
The latter class of sources is variously termed 
as ``optically-dull", ``optically normal" or \xbongs\  (X-ray Bright Optically 
Normal Galaxies; Comastri et al. 2002). The common meaning of these definitions is that 
they lack evidence of accretion-driven activity in their optical spectra, 
in contrast with ``normal" Seyfert galaxies and quasars. 
Their X-ray luminosities ($\approx 10^{42}-10^{43}$ \lum), X-ray spectral shape and  
X-ray-to-optical flux ratio (X/O\footnote{Where X/O is defined as 
$X/O=\log{\frac{f_X}{f_R}}=\log{f_X}+C+\frac{R}{2.5}$.}$\sim -1$) 
suggest AGN activity of  moderate strength.   
Originally discovered  in early {\it Einstein} observations (Elvis et al. 1981) 
and named optically dull galaxies, the interest on the nature of these sources 
has gained a renewed attention after the discovery of 
several examples in XMM-{\it Newton} and {\it Chandra} surveys
(Fiore et al. 2000; Comastri et al. 2002a,b; Georgantopoulos et al. 2005; Kim et al. 2006). 
Several possibilities were proposed 
in the literature in order to explain why a relatively luminous, hard X-ray source 
does not leave any significant signature of its presence in the 
form of emission lines.

A simple explanation favoured by Moran et al. (2002) and more recently 
by Caccianiga et al. (2007) for faint sources in the 
\chandra\ deep fields and brighter object in the \xmm\ XBS survey, respectively, is 
dilution by the host galaxy starlight. 
The combination of optical faintness and lack 
of strong emission lines in the observed wavelength range for distant \chandra\
sources  or the inadequate observing set-up 
among brighter nearby \xmm\ objects (Severgnini et al. 2003) may account 
for the \xbong\ properties. 
More in general, if the contrast between the host galaxy starlight and nuclear emission 
is high, AGN emission lines may easily be undetected. 
The physical reason may be ascribed to obscuration, most likely with a large covering factor, 
or to the fact that the lines are not efficiently produced by the central engine.

%However, it has been shown that this is not the case 
%for a significant fraction of  ``local'' \xbongs\ on the basis      
%of the upper limits on the {\rm O[III]$_{5007\AA}$} luminosity (Cocchia et al. 2007) 
%as well as for more distant z $\sim$ 1 sources, for which the optical-to-X-ray 
%luminosity ratio is fully consistent with that of ``normal" Seyferts and quasars
%(Rigby et al. 2006).
%
%For those sources for which the dilution hypothesis can be safely ruled out, the alternative 
%possibilities could be grouped in two classes. The first one is that \xbongs\ are merely obscured AGN:
If \xbongs\ are merely obscured AGN, two hypotheses may be envisaged:

\par
$\bullet$ In order to explain 
the multiwavelength  properties of the \xbong\ prototype PKS~0312018, 
also known as P3, Comastri et al. (2002) suggested 
heavy obscuration by Compton-thick gas
covering almost 4$\pi$ at the nuclear X-ray sources. In this way, no 
ionizing photons can escape to produce the narrow emission lines 
which are observed in ``normal" Type 2 narrow-line AGN 
which are thought to have a lower 
covering fraction following the AGN Unified Scheme (but see Section \ref{xray} 
for a recent re-analysis of the X-ray data of P3).  

\par
$\bullet$
According to a detailed multiwavelength analysis of  ``optically-dull" galaxies
in the \chandra\ deep fields, Rigby et al. (2006) conclude that extranuclear 
dust in the host galaxy plays an important role in hiding the emission lines.\\
 
Alternatively, \xbongs\ may be members of a class, or classes, of {\it exotic} 
objects for which emission lines are either intrisically weak or absent:

\par
$\bullet$ Radiatively Inefficient Accretion Flows (RIAFs) are expected at 
accretion rates well below those inferred for Seyferts and quasars. 
A distinctive property of low accretion-rate flows is that the standard 
Shakura-Sunyaev accretion disk is truncated at a relatively large inner radius.
As a consequence, it cannot generate the ``big blue bump" and enough UV photons
to photoionize the line-emitting circumnuclear gas. The infalling gas is heated 
to high temperatures and emits a hard X-ray power-law by upscattering 
of low-energy seed photons. According to Yuan \& Narayan (2004), the SED of source P3 
could be reproduced by a RIAF model. 

\par
$\bullet$ 
\xbongs\ could be extreme BL Lac objects in which the featureless 
non-thermal continuum is much weaker than the host galaxy starlight. 
Following Fossati at al. (1998), \xbongs\ could belong to the 
low-luminosity tail of the blazar spectral sequence based on the anti-correlation 
between luminosity and frequency of the synchrotron peak.

\par
$\bullet$ 
A highly speculative possibility is that of a transient AGN phenomenon in the process
of tidally disrupting a star. If this were the case, the X-ray emission should be 
witnessing the transient accretion phenomenon (see Komossa et al. 2004
for extreme variability events in ROSAT observations; see also Gezari et al. 2006 
for a luminous flare observed in the GALEX Deep Imaging Survey). 
The transient is most likely over in subsequent follow-up optical observations.

Finally, it should be noted that diffuse emission from a galaxy group, whose
X-ray extended emission may have escaped detection in low signal-to-noise
X-ray observations, is also possible and indeed observed in a few cases 
(Georgantopoulos et al. 2005).

%Finally the absence of the emission lines might be related to 
%the fact that \xbongs\ might be gas poor galaxies or the gas covering factor 
%is much lower than in Seyfert 2 galaxies. \\

%Although some of the explanations mentioned above provide a good
%description of the observed properties for a few objects, the very nature
%of \xbongs\ is still subject of debate. 

While a unique solution may not necessarily hold for all the \xbongs\
observed in different surveys, 
they represent a useful benchmark for a better understanding 
of the AGN activity and, as such, deserve further studies.
Ideally, one would need sensitive, high-spatial resolution, 
multiwavelength observations from radio to X-rays. 
As a first step, in the following we use good-quality 
photometric near-infrared 
data obtained with ISAAC at VLT of four low-redshift ($z=0.1-0.3$) \xbongs, selected in the 
{\rm HELLAS2XMM} survey to search for the presence of a putative nucleus 
which has escaped detection in the optical spectroscopy. 
The rather obvious advantage of near-infrared data is that 
the effects of dust reddening are minimized and that the 
nuclear emission in this 
band should rise more rapidly than the stellar light due to the reprocessing by hot dust. 
At the same time, the excellent 
quality of the near-IR images makes possible 
to apply a surface brightness decomposition technique, already successfully applied 
by several authors (e.g. S\'anchez et al. 2004; Peng et al. 2006), to
search for weak unresolved nuclear emission down to faint near-infrared magnitudes. 
We also discuss the broad-band properties of the four \xbongs\ using 
available multiwavelength data.  
Throughout the paper we assume a cosmological model
with $H_0$ = 70 km s$^{-1}$ Mpc$^{-1}$, $\Omega_m$ = 0.3 and
$\Omega_{\Lambda}= 0.7$.

\section{The sample}

In the following, the term \xbong\ is referred to relatively nearby ($z <$ 0.3), low luminosity
($L_X \sim 10^{42-43}$ erg s$^{-1}$), hard X--ray (2--10 keV) selected sources with an optical spectrum
typical of early type galaxies. All together they represent a small fraction ($\sim$ 5\%) of 
the sources in hard X--ray surveys.
The four \xbongs\  were selected from the 
HELLAS2XMM survey: 3 sources belong to the one square degree field (1dF) sample 
(see Baldi et al. 2002 for a description of the 
X--ray data and Fiore et al. 2003 for the optical identification process), while one 
object (Abell 1835140) is from the survey extension (Cocchia et al. 2007). 
They represent 40\% of the HELLAS2XMM \xbong\ sample (10 objects), and have been selected for  
infrared observations on the basis of a high quality optical 
spectrum and bright optical magnitudes (R$<$18).

%The multiwavelength analysis of the \xbong\  prototype 
%PKS~03120018 (also known as P3) is presented in Comastri et al. (2002).  
%All four sources have hard (2--10 keV) X-ray unabsorbed luminosities 
%larger than $\sim 10^{42}$ erg s$^{-1}$, while the optical spectra are characterized 
%by a red continuum with stellar absorption lines (Fig. \ref{optspectra}). 
\begin{figure}
\centering
\includegraphics[width=0.8\textwidth]{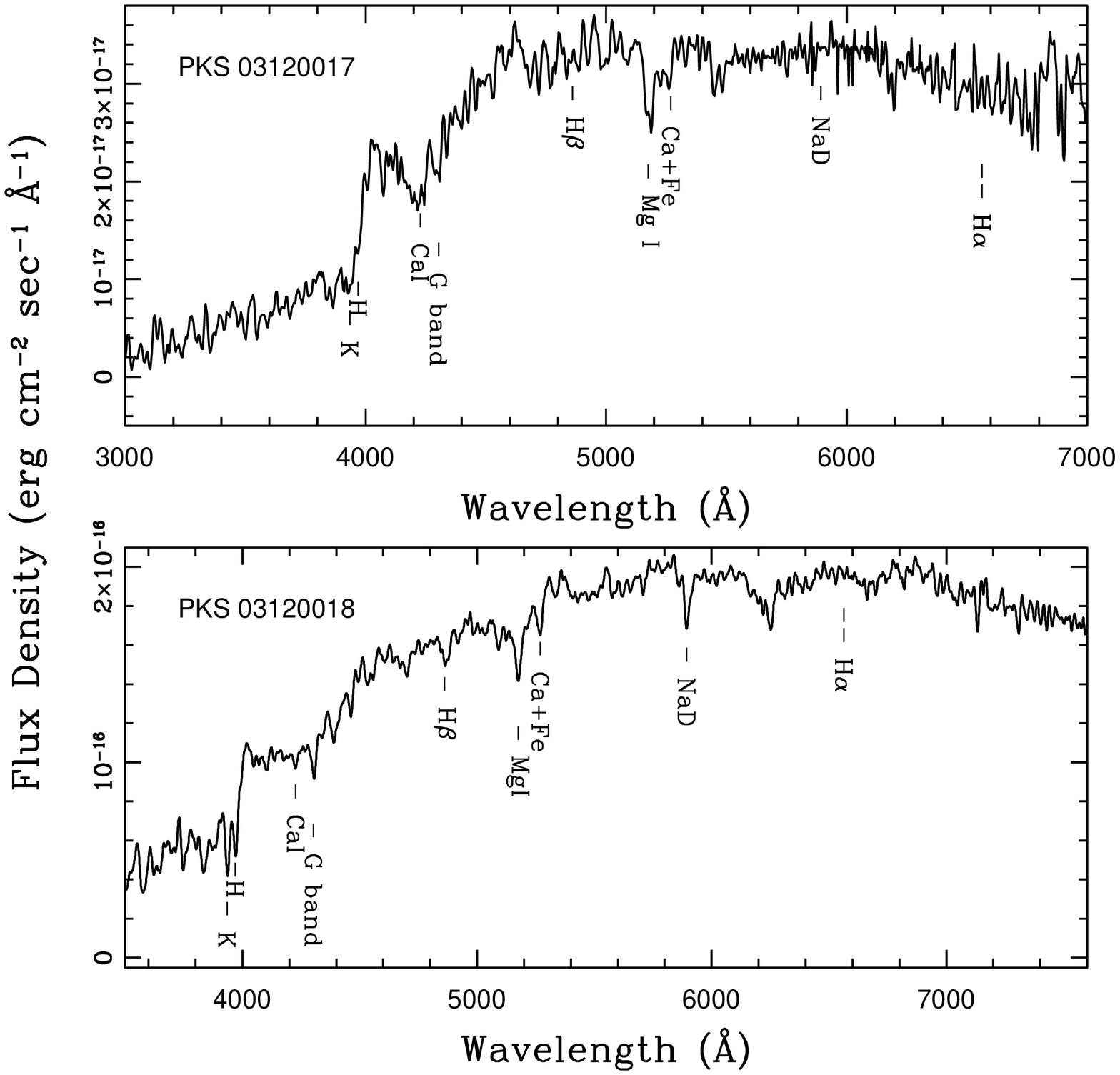}
\includegraphics[width=0.8\textwidth]{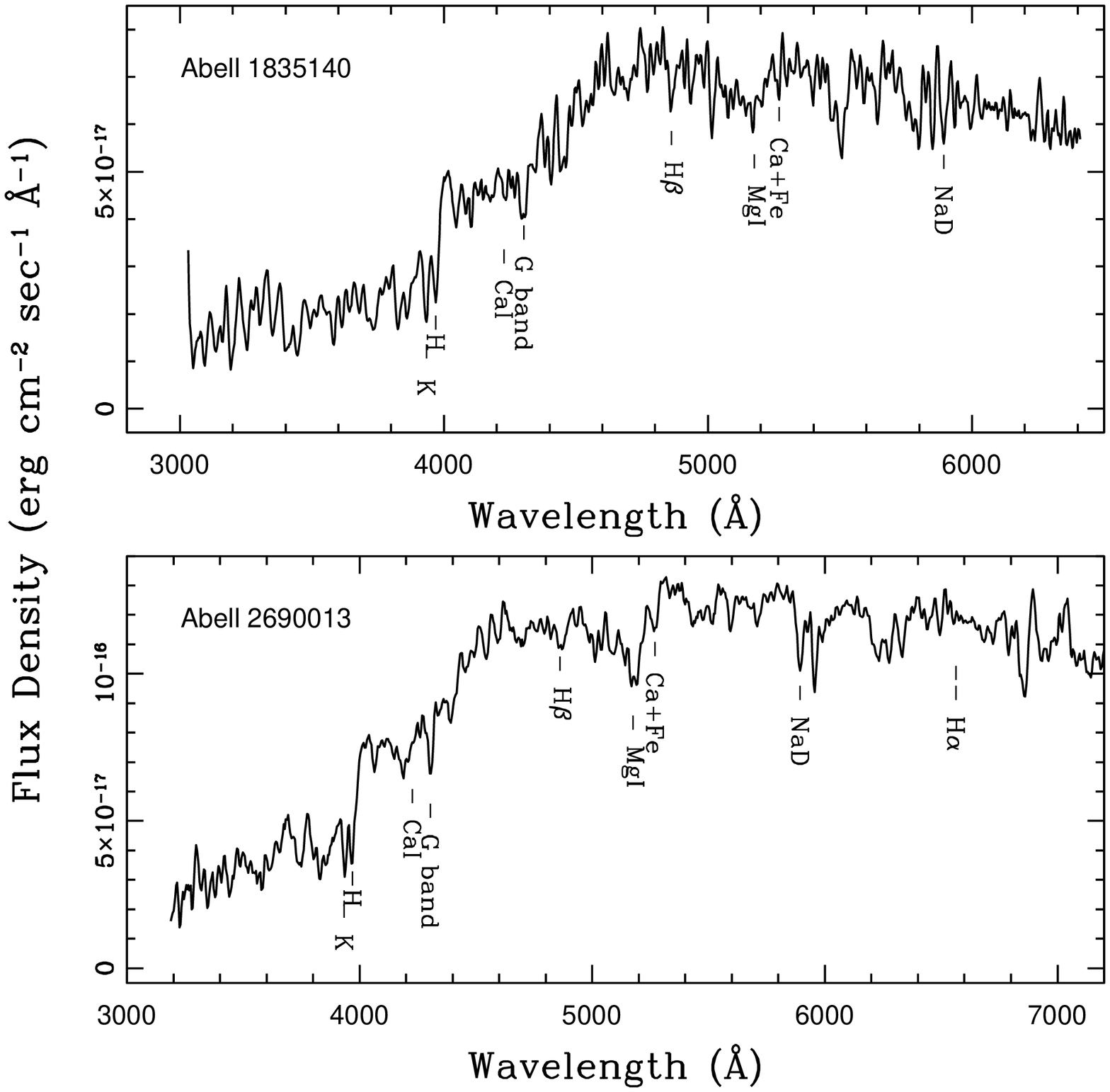}
\caption{Rest-frame optical spectra of the sources PKS 03120017 and PKS 03120018, Abell 2690013 and Abell 1835140, taken with the ESO 3.6m telescope. The stellar absorption lines are labeled. The H$\alpha$ line is also labeled at the expected wavelength.}
\label{optspectra}
\end{figure}

The spectroscopic observations were obtained at the ESO 3.6m telescope
equipped with EFOSC2 during different observing runs in 2001. 
The optical spectra, all of them characterized by absorption features
over a red stellar continuum, are shown in Fig. \ref{optspectra}.
While the relatively large slit width (2 arcsec) used in the spectroscopic observations 
do not preserve us from a galaxy light contamination (and consequent line dilution), 
the good signal-to-noise ratio 
(S/N$_{6000\AA} >$13--20) and the well suited spectral range (including the expected wavelength 
of the $[OIII]_{5007\AA}$ and $H\alpha_{6563\AA}$ lines, with the exception of Abell~1835140) allow us 
to place firm constraints on the line fluxes.  

\begin{figure}
\centering
\includegraphics[width=0.6\textwidth]{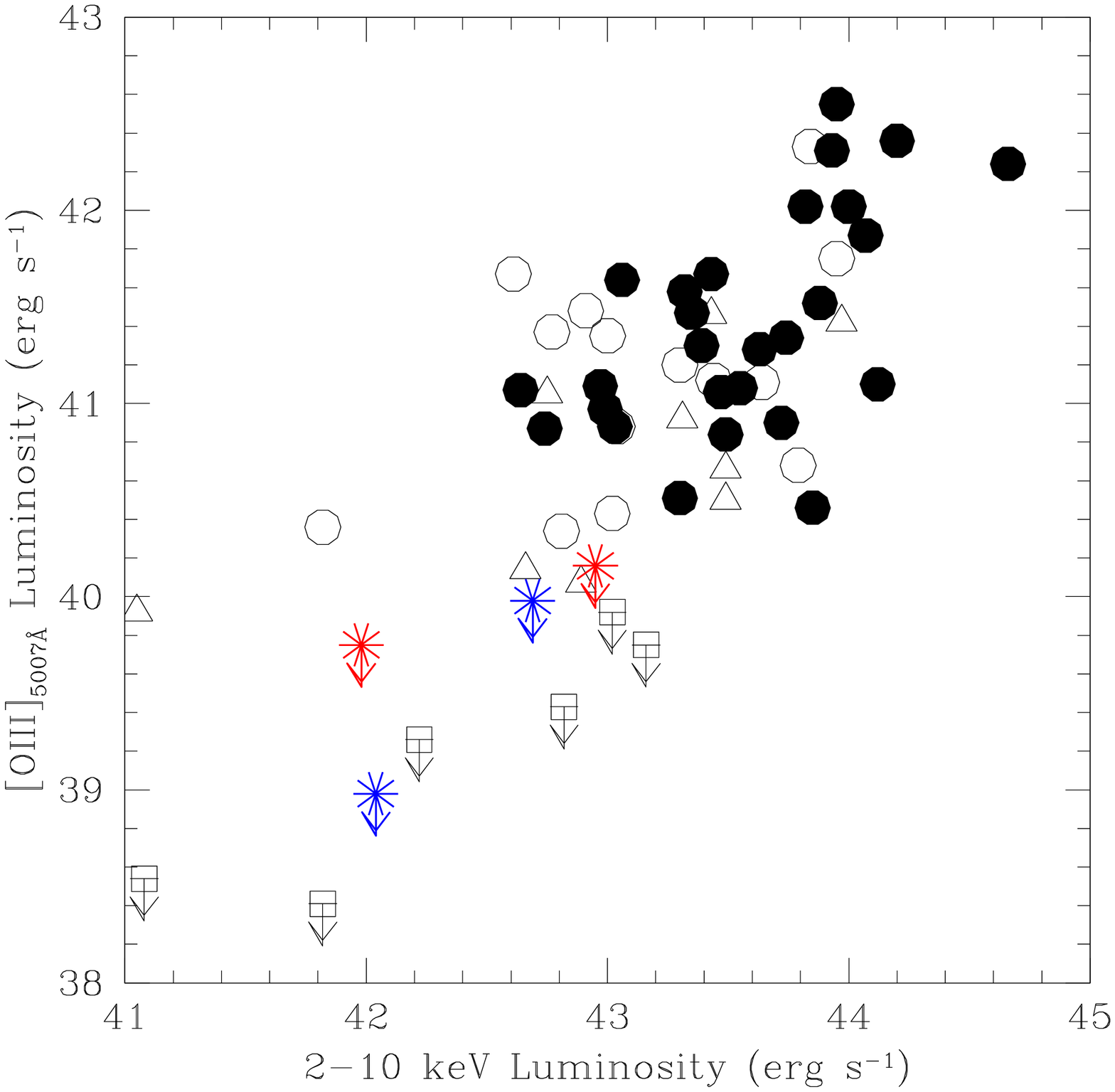}
\caption{The $[OIII]_{5007\AA}$ luminosity  versus the 2--10 keV luminosity for 59 sources of the {\rm HELLAS2XMM} 
sample as presented in Cocchia et al. (2007). Starred symbols mark the four \xbongs\ discussed in this paper; 
filled circles = broad-line AGN; open circles = narrow-line AGN; open triangles and open squares = emission-line and early-type galaxies, respectively. }
\label{lx_loIII}
\end{figure}

The corresponding luminosity upper limits for the 
{\rm O[III]} line of the four \xbongs\ (starred symbols in Fig. \ref{lx_loIII})
are in the range 10$^{39}$--10$^{40}$ erg s$^{-1}$, about 
one order of magnitude fainter than that measured in the optical spectra
of {\rm HELLAS2XMM} broad-line (BLAGN, filled circles) and narrow-line (NLAGN, open circles) AGN with comparable X-ray luminosity 
(see Fig. \ref{lx_loIII} and Cocchia et al. 2007). 
The upper limits are more than one order of magnitude below the relation between 
X--ray luminosity and {\rm O[III]} to X--ray luminosity ratio obtained by Netzer 
et al. (2006) for a large sample of X--ray selected Type 1 and Type 2 AGN spanning
the same range of X--ray luminosities.
We also note that only 0.4\% of DR6 SDSS AGN (about 7000 quasars; Adelman-McCarthy et al. 2007) 
have {\rm O[III]} luminosity below 10$^{40}$ erg s$^{-1}$.

The redshifts obtained from the position of the principal absorption lines are 
$z~=~0.319,~0.159,~0.154,~0.251$ for PKS~03120017, 
PKS~03120018, Abell~2690013 and Abell~1835140, respectively (Fiore et al. 2003; Cocchia et al. 2007). 
The X-ray unabsorbed luminosities are reported in Table \ref{tabxray}.
Radio observations at 5 GHz are available for the two objects in the PKS 0312-77 field. 
Source PKS~03120017 was detected with a flux density of 1.3 mJy at 5 GHz, while 
source PKS~03120018 remains undetected (F$_{5GHz} <$ 0.15 mJy at the 3$\sigma$ level). An observations at 8.7 GHz 
is also available for source PKS~03120017 which was detected with a flux density of 0.5 mJy. 

\section{X-ray spectral analysis}
\label{xray}
The results of the spectral analysis of the X-ray sources in the 1dF sample
are presented in Perola et al. (2004) with particular emphasis towards the statistical
properties of the sample. 
The X-ray spectra of the four \xbongs\ discussed 
here were re-analyzed including an 
additional observation of the Abell~1835140 field (not included in the 
Perola et al. paper) using the most recent calibrations.
The data were processed using standard {\sc sas\footnote{\xmm\ Science Analysis System; see 
http://xmm.vilspa.esa.es/external/xmm\_sw\_cal/sas\_frame.shtml.} v6.1.0} and {\sc ftools} tasks. 

\begin{figure}
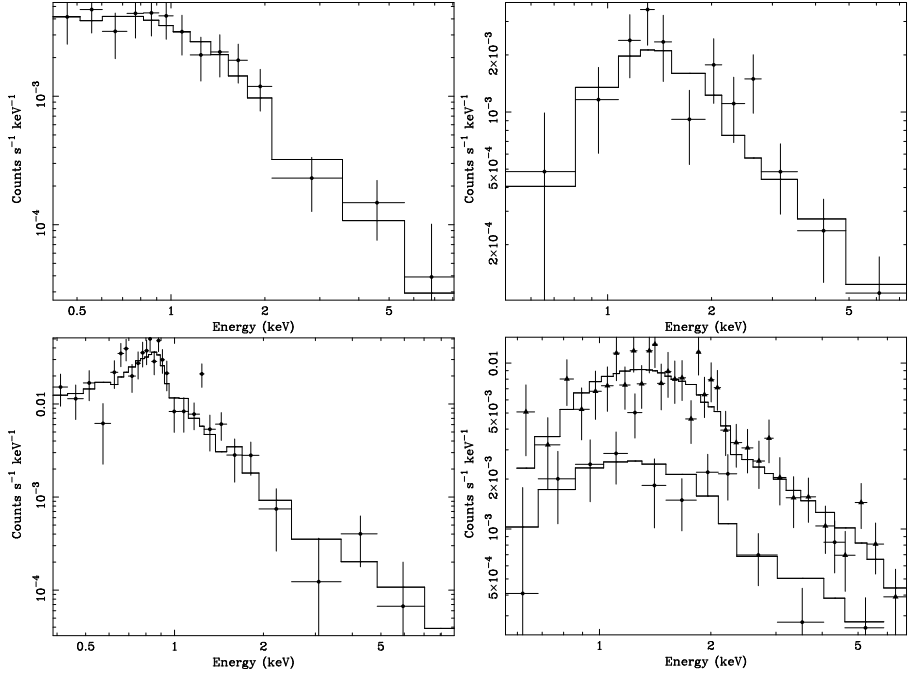

\centering
\includegraphics[scale=0.25,angle=-90]{civano_3_1.ps}
\includegraphics[scale=0.25,angle=-90]{civano_3_2.ps}
\includegraphics[scale=0.25,angle=-90]{civano_3_3.ps}
\includegraphics[scale=0.25,angle=-90]{civano_3_4.ps}
\caption{XMM-Newton (pn) spectra of the sources PKS~03120017, PKS~03120018, 
Abell~2690013 and Abell~1835140 (from top-left to bottom-right), fitted in the 0.5--8 keV range. 
The solid line represents the best-fitting model. The spectral parameters are reported in Table ~\ref{tabxray}. For source Abell~1835140, the spectra of the two available observations are reported. }
\label{xray-spec}
\end{figure}

The pn spectra were grouped with a minimum of 10 counts per bin (for PKS~03120018 
and PKS~03120017) and 15 counts per bin (for Abell~26900013 and Abell~1835140) 
using the task {\sc grppha}, and fitted with {\sc xspec v11.3.0} (Arnaud 1996), 
while MOS1 and MOS2 data, having much lower counting statistics, 
were left unbinned and fitted with the C-statistic (Cash 1979).
We checked and found consistency between the pn and MOS data. However,  
the X-ray spectral results reported in Table~\ref{tabxray} and the best-fitting spectra shown in 
Fig. \ref{xray-spec} are referred to the pn data alone, given the lower counting statistics of the MOS data. 
%%%%
\begin{table}[b]
\centering
\footnotesize
\caption[]{X-ray spectral analysis parameters of the pn data. The X-ray luminosities are corrected for absorption. For source Abell~1835140, the spectral parameters of the two available observations are reported.}
\begin{tabular}{ccccccc}

\hline
\hline
Src.~ID & kT    & $\Gamma$ & $N_H$       & $\chi^2$/d.o.f & F$_{2-10~keV}$             & $L_{2-10~keV}$   \\
        & (keV) &          & (cm$^{-2}$) &                & (erg~cm$^{-2}$~s$^{-1}$)   & (erg~s$^{-1}$)   \\
\hline
PKS~03120017   &                 & 2.4$^{+0.6}_{-0.4}$ & $<2.1\times10^{21}$                & 3.7/11  & 1.4$\times10^{-14}$ & 5.0$\times10^{42}$ \\
PKS~03120018   &                 & 2.1$^{+0.8}_{-0.4}$ & 6.9$^{+7.9}_{-4.6}\times10^{21}$ & 9.6/9   & 1.5$\times10^{-14}$ & 1.1$\times10^{42}$ \\
Abell~26900013 & 0.73$\pm{0.06}$ & 1.6$^{+0.6}_{-0.9}$ &                                     & 34.4/25 & 1.5$\times10^{-14}$ & 9.6$\times10^{41}$ \\
Abell~1835140 &                 & 1.5$^{+0.6}_{-0.5}$ & 4.3$^{+5.0}_{-2.4}\times10^{21}$ & 14.0/12 & 5.3$\times10^{-14}$ & 9.0$\times10^{42}$ \\
               &                 & 1.7$^{+0.3}_{-0.3}$ & 6.5$^{+2.9}_{-2.2}\times10^{21}$ & 32.9/35 & 1.8$\times10^{-13}$ & 3.3$\times10^{43}$ \\
\hline
\hline
\label{tabxray}
\end{tabular}
\end{table}
%%%
In the following we summarize the main X-ray spectral results.\\
Two sources, PKS~03120018 and Abell~1835140, are best fitted by 
an absorbed power law with photon index in the range $\Gamma \approx$ 
1.5--2.1 and column densities ($N_H \approx~4-7\times10^{21}$~cm$^{-2}$), typical 
of a mildly obscured AGN. Despite the relatively large errors in the determination 
of the X-ray spectral parameters, this analysis seems to rule out
Compton-thick absorption, at least in the energy range covered by \xmm. 
The radio-loud 
source PKS~03120017 is best fitted by a relatively steep $\Gamma \simeq$ 2.4 spectrum (but consistent with $\Gamma \sim$ 2)
and no absorption in excess to the Galactic value (see Table~\ref{tabxray}).\\

\begin{figure*}
\centering
\includegraphics[height=7.5cm,width=10cm]{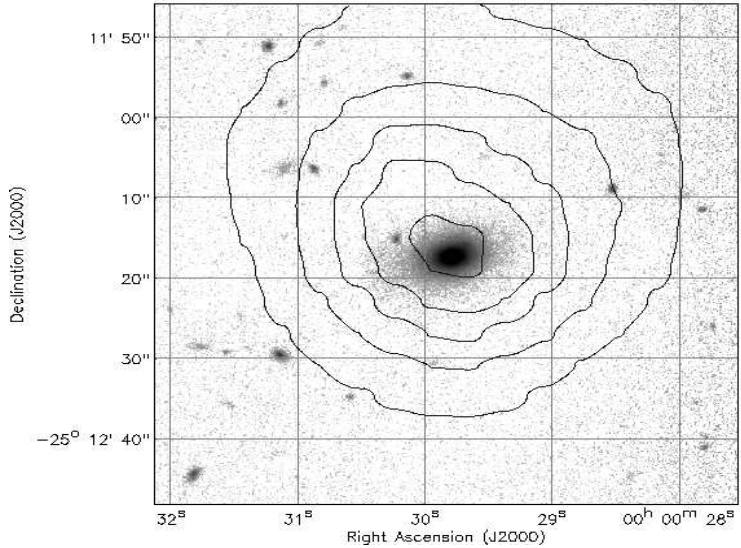}
\caption{X-ray contours obtained
 from the 0.5--10 keV adaptively smoothed pn image of source Abell~2690013 overlaid on the \kei band image. The image is $\approx$ 60''$\times$60''.}
\label{contorni}
\end{figure*}

Abell~26900013 is well fitted using a thermal ({\sc mekal}) model at soft 
energies and a weakly constrained power-law tail at higher ($\gtrsim$2~keV) energies. 
The thermal spectrum is consistent with the extended appearance of the X-ray emission, 
as shown by the overlay of the X-ray contours on the \kei band image (see Fig.~\ref{contorni}). \\

Significant X-ray flux variability (a factor about 3) is present between 
the two \xmm\ observations of  Abell~1835140. This object 
was also observed twice and detected by \chandra\ 
(in 1999 and 2000, for $\approx$~20 and 10~ks, respectively).
%Unfortunately, it was at very large off-axis radii, making it impossible 
%to deconvolve the X-ray emission from the two optical close nuclei (less than 1 arcsec, see below). 
The \chandra\ data (especially the data of the first and longer observation) confirmed 
the \xmm\ spectral results and allowed for a long-term variability study of this source. 
Over $\approx$~3 and half years, Abell~1835140 was found to be variable by a factor of 
$\approx$~7.5, with its 2--10~keV flux increasing from the value of 
$\approx2.4\times10^{-14}$\cgs\ in 1999 to $\approx1.8\times10^{-13}$\cgs\ in 2003, 
with no evidence for significant changes in the X-ray spectral parameters. Even if the flux variability could be connected with the tidal disruption of a star, the increasing of the X-ray flux 
allows us to rule out such a possibility.

\section{Near-Infrared data analysis}

\subsection{Near-Infrared imaging observations}

\begin{figure}
\centering
\includegraphics[angle=0,width=0.53\textwidth]{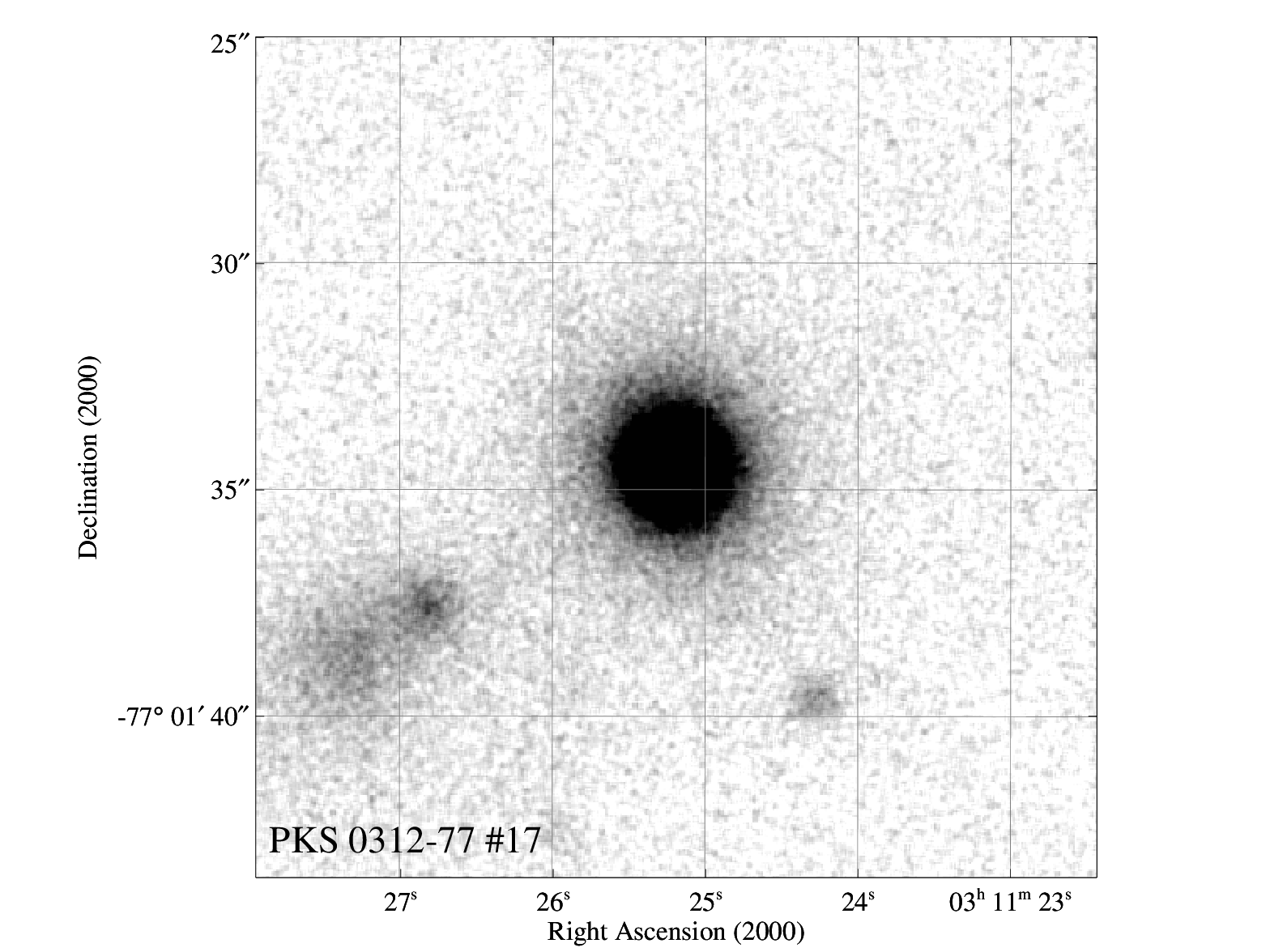}
\vglue0.3cm
\includegraphics[angle=0,width=0.53\textwidth]{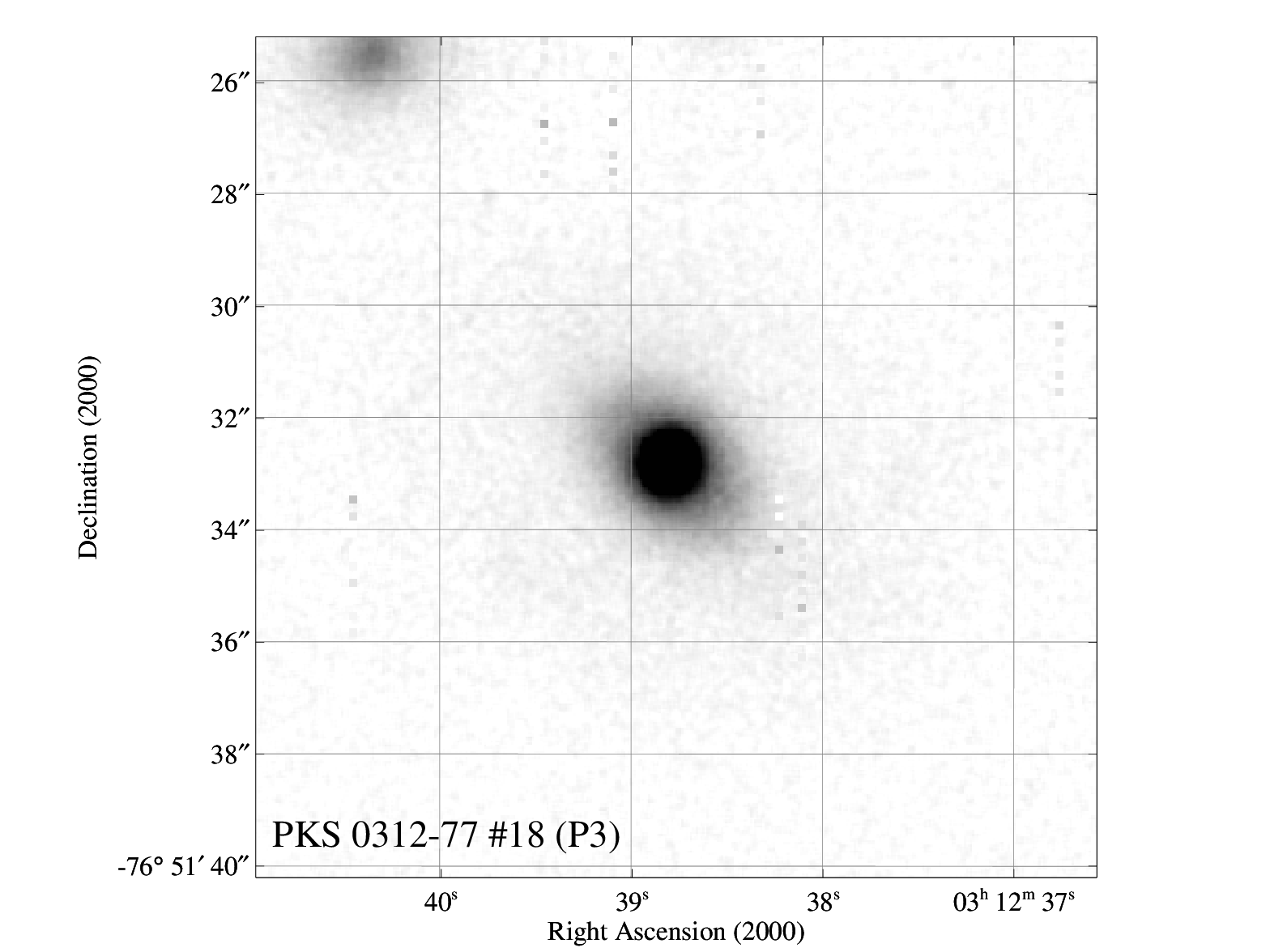}
\hglue-1.2cm
\includegraphics[angle=0,width=0.53\textwidth]{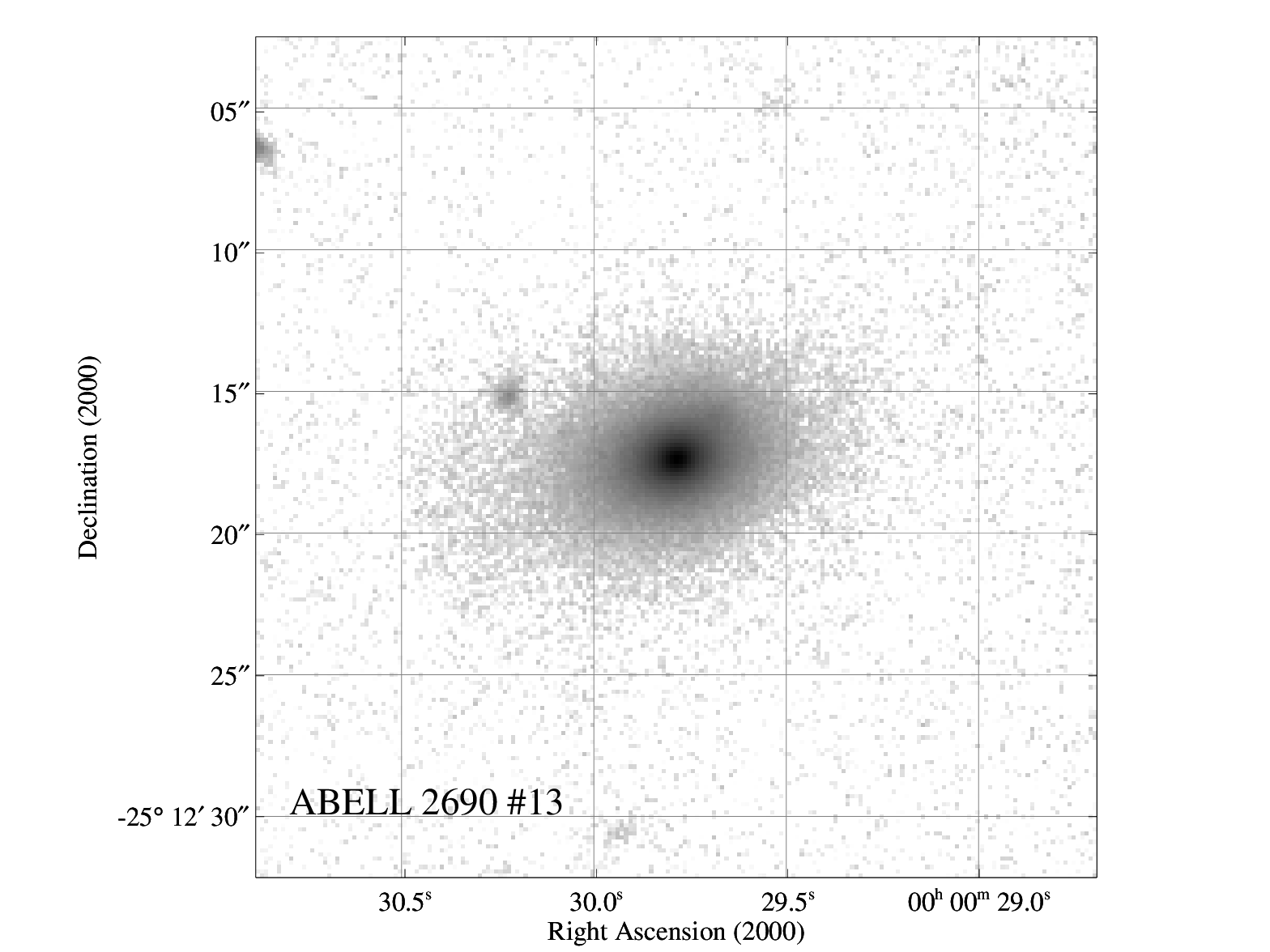}
\hglue-1.2cm
\includegraphics[angle=0,width=0.53\textwidth]{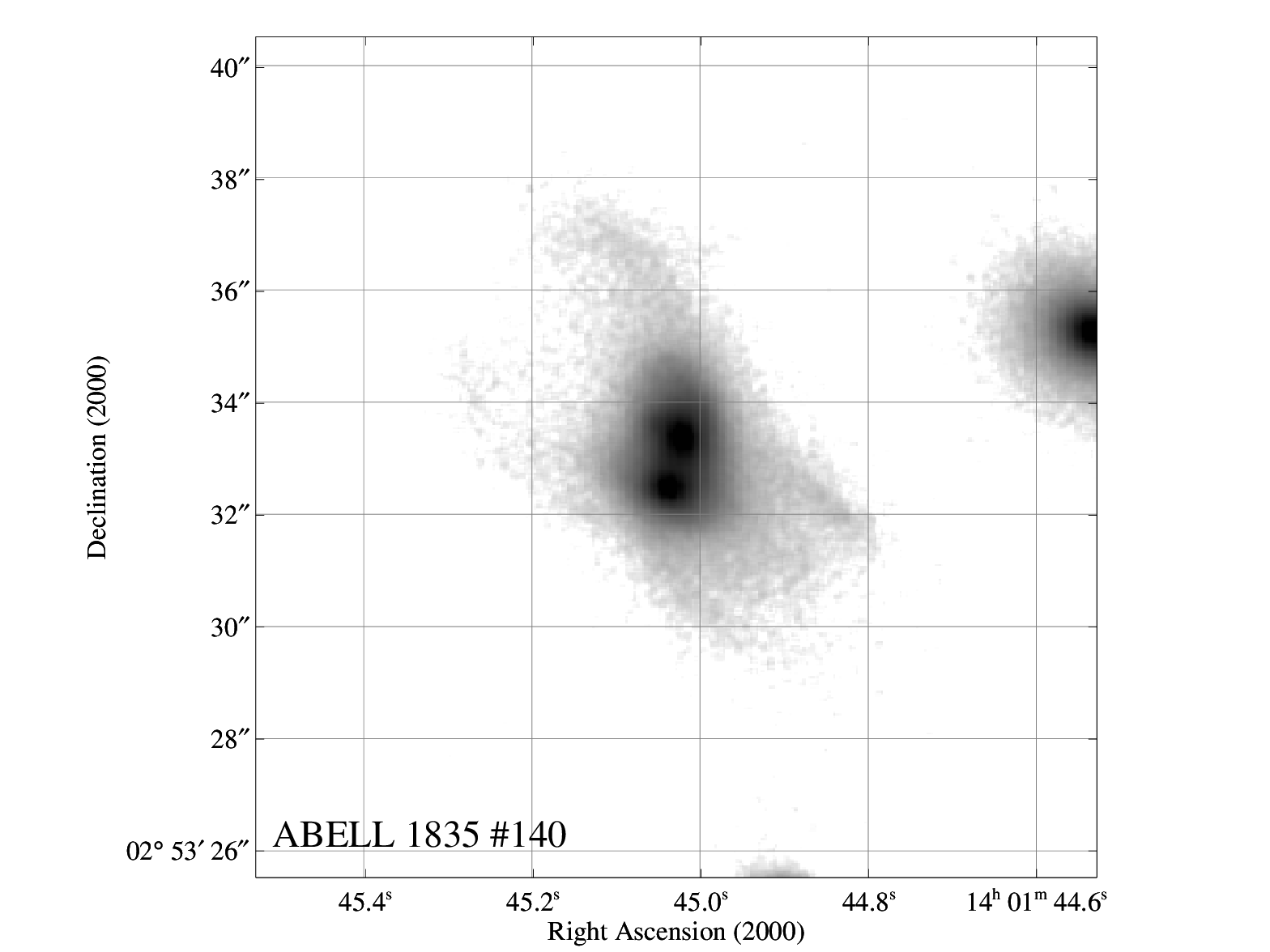}
\caption{\kei band images: PKS 03120017, PKS 03120018, Abell 2690013 and Abell 1835140, from top left to bottom right; the size of the images 
is 20''$\times$20'', 15''$\times$15",  30''$\times$30", 15''$\times$15", respectively.}
\label{xbong_k}
\end{figure}

Deep near-infrared observations 
have been carried out using the Infrared Spectrometer and Array Camera ($ISAAC$) mounted 
on the ESO VLT (Very Large Telescope).\ $ISAAC$ is equipped with 
a Rockwell Hawaii 1024$\times$1024 HgCdTe array, with a pixel scale of 
0\farcs147/pixel and a field of view of 2\farcm5 $\times$ 2\farcm5. 
The observations have been collected in service mode over several nights 
during September 2002 under good seeing conditions ($<$0\farcs8).
The ISAAC images were taken in two near-IR bands 
(the \GEI\ and \KEI\footnote{The \KEI\ (K-short) filter is centered at shorter wavelenght 
than the standard $K$ filter in order to reduce thermal background.}\ filters)
and were reduced using the DIMSUM\footnote{Deep Infrared Mosaicing Software,
developed by P.Eisenhardt, M.Dickinson, A.Stanford and J.Ward.,
and available at the site {\tt ftp://iraf.noao.edu/contrib/dimsumV2}.}
package, following standard procedures. A detailed discussion on the data reduction of 
similar images taken for other sources revealed in the {\rm HELLAS2XMM} 1dF is reported 
in Mignoli et al. (2004). The total exposure 
times for each galaxy is 600 sec in \GEI\ and 1800 sec in \KEI\ band.
The effective seeing on final frames ranges between 0.5 and 0.8 arcsec (FWHM). 
The magnitude limits for 
point-like sources are about \GEI(lim)=22 and \KEI(lim)=21.5 in all the images. The \kei band images 
of all the four sources are shown in Figure \ref{xbong_k}.

% the limits are for a true background
\subsection{The fitting procedure}
In order to search for the presence of a putative nucleus, we applied the surface brightness decomposition 
technique using GALFIT (Version 2.0.3b, Peng et al. 2002), 
a two-dimensional algorithm designed to extract structural parameters directly from 
galaxy images combining
several analytical models and by convolving them with the image Point 
Spread Function (PSF). 
The PSF has been modeled 
adopting several field stars in each ISAAC image. The fitting has been performed in a box around each object 
large enough (220$\times$220 pixels) to contain most of the bulge.

The fitting procedure described below was performed separately in both \GEI\ and \KEI\ bands, allowing us 
to cross-check the results 
and test for a possible color trend. In order to better constrain the galaxy structural 
parameters (magnitude and effective radius), first we fitted each object in the \GEI\ band, 
where the AGN contribution to the total light was expected to be lower 
(according to a standard AGN SED), as actually found in our analysis. 
Then we applied GALFIT to the \kei band images, using the galaxy 
parameters obtained in the \GEI\ band as start-up values 
for the nuclear decomposition.  

As a first guess of the profile of the host galaxy, we applied a 
S\'ersic model (S\'ersic 1968) of the form:
  
\begin{equation} \mu(r) = \mu_e \mbox{ e}^{-\kappa \left[({\frac{r}{r_e}})^{1/n} - 
1\right]} \end{equation}

where $\mu_e$ is the effective surface brightness, $r_e$ the effective radius, 
$n$ the S\'ersic index, and $\kappa$ a constant determined from $n$ in order to have 
half of the integrated flux within $r_e$. 
The S\'ersic index assumes the value $n \sim1$ in case of exponential profile and value $n=4$ in case of a 
de Vaucouleurs profile (de Vaucouleurs 1948).
A fit with a S\'ersic model returns $n \gtrsim $ 6 in all the sources.
Since this index is higher than the typical 
values found in early-type galaxies (e.g. Andredakis et al. 1995; Cassata et al. 2005) and given the 
early type like galaxy spectra of our sources, we adopted a de~Vaucouleurs profile, hereafter referred to as MODEL~1.

If residual emission in the innermost region were still present after fitting the galaxy profile with MODEL~1, 
we added the contribution of a point-like source (MODEL~2). 
The latter is modeled using the PSF profile obtained by averaging stars in the field.
The centroid of the unresolved component and that of the host galaxy have been 
linked each other in the fitting procedure. 
We considered the unresolved component as detected if (1) the $\chi^2$ significantly improves with 
the inclusion of the nuclear component (see Section \ref{fr}) and (2) it is brighter than 
the image magnitude limit for point-like sources. The results of the fitting are shown in Table~\ref{galfit}. 

In order to check the GALFIT capabilities to detect 
a faint unresolved source at the center of a bright galaxy, 
accurate simulations were performed using the IRAF task \textit{mkobject}. 
We started with a $r^{1/4}$ profile and then a point-like source has been added. 
In the simulations, we adopted a grid of magnitudes in the range \KEI=17 -- 22 for the point-like 
component and \KEI=14 -- 22 for the host galaxy; different effective radii (from 30 to 120 pixels) 
for the host component have been used.
The input parameters have been chosen in an appropriate range to reproduce the observed data 
and to span a wider but realistic parameter space. 
The simulated images have been convolved with the seeing and the noise of our images; 
the PSF has been modeled with the parameters determined from the field stars. 
We applied GALFIT on the simulated sources first modeling with a de Vaucouleurs profile and 
then adding a central unresolved component, as performed for the \xbongs.  
The fitting results are fully consistent with the input parameters. Uncertainties ($\Delta\ mag \sim 0.02$) 
on the determination of the nuclear magnitude have been found in the case of a very faint (\KEI\ $\gtrsim$22.5) 
unresolved source hosted by a bright (\KEI\ $\simeq$15) galaxy. 
We conclude that at the magnitude limit of our images
for point-like sources, the errors on the parameters due to the GALFIT detection capabilities 
are not affecting our results significantly.

\subsection{The fitting results}
\label{fr}
\begin{table}[t]
\centering
\footnotesize
\caption{Results of the fits performed with GALFIT in the \GEI\ and in the \KEI\ band; 
the two entries for each source in each band correspond to the fitting with MODEL 1 and with MODEL 2, respectively.}
\begin{tabular}{ccccccccc}
\hline
\hline
id & band & Model& m$_{host}$ & r$_{eff}$ (kpc)  & m$_{nucleus}$ & $\chi^2$& d.o.f.&$\chi^2/ d.o.f$\\
\hline
PKS03120017 & \GEI & 1 & 16.94$\pm$0.03  &4.79$\pm$0.51    & & 45983& 48835&0.94   \\
	    & \GEI & 2 & 16.92$\pm$0.05  &5.64$\pm$0.86     &20.75$\pm$0.29  &45614&48832 & 0.93   \\
    	    & \KEI & 1 & 15.24$\pm$0.01  &4.96$\pm$0.17     & & 54680& 48835& 1.12   \\
	    & \KEI & 2 & 15.21$\pm$0.01  &6.50$\pm$0.10     &18.39$\pm$0.24    & 49722& 48832& 1.02   \\ 

PKS03120018 & \GEI & 1  &16.51$\pm$0.02& 3.73$\pm$0.30  &  &75401& 48835&1.54\\
	    & \GEI & 2  &16.52$\pm$0.01& 4.03$\pm$0.40 &  20.30$\pm$0.35 & 67046 & 48432& 1.37\\
	    & \KEI & 1  &15.25$\pm$0.01 & 3.30$\pm$0.05  & &85728&48835& 1.75\\
 	    & \KEI & 2  &15.20$\pm$0.01 & 5.04$\pm$0.20  &18.09$\pm$0.10& 71944&48832&1.47\\

Abell2690013  & \GEI & 1  &15.41$\pm$0.01 & 7.34$\pm$0.10  & &51227&48835& 1.05   \\
	      & \KEI & 1  & 14.29$\pm$0.01  &6.86$\pm$0.20  & & 106194& 48835& 2.17 \\
	        
\hline
\hline
\label{galfit}

\end{tabular}

\end{table}

The best-fitting parameters (m$_{host}$, radius, m$_{nucleus}$) obtained in each 
band for MODEL~1 and MODEL~2 are reported in Table~\ref{galfit}. 
The fifth column of Table~\ref{galfit} gives the effective radius, which is always larger than the seeing disk
(up to 3.5 kpc at z = 0.3).
Since the uncertainties quoted by GALFIT are very small (see Peng et al. 2002), 
the errors reported in Table~\ref{galfit} have been estimated comparing the best-fitting results obtained 
using the different stars in the images one by one as a model for the PSF; this method has been 
applied because it is known that steep profiles are very sensitive to the accuracy of the used PSF 
(see the online manual\footnote{http://zwicky.as.arizona.edu/~cyp/work/galfit/TFAQ.html.}). 

\begin{figure*}
\centering
\includegraphics[width=0.7\textwidth]{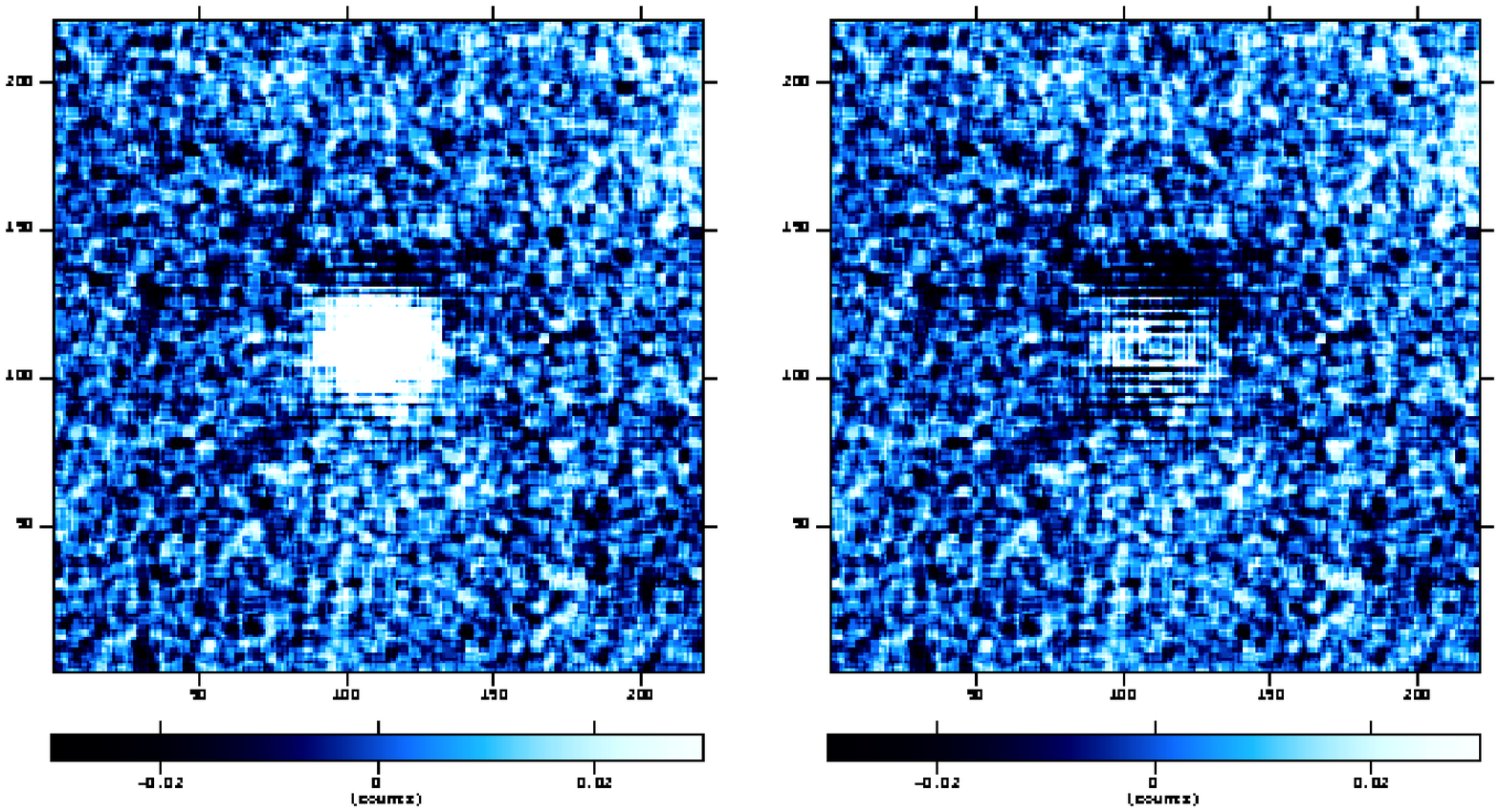}
\includegraphics[width=0.7\textwidth]{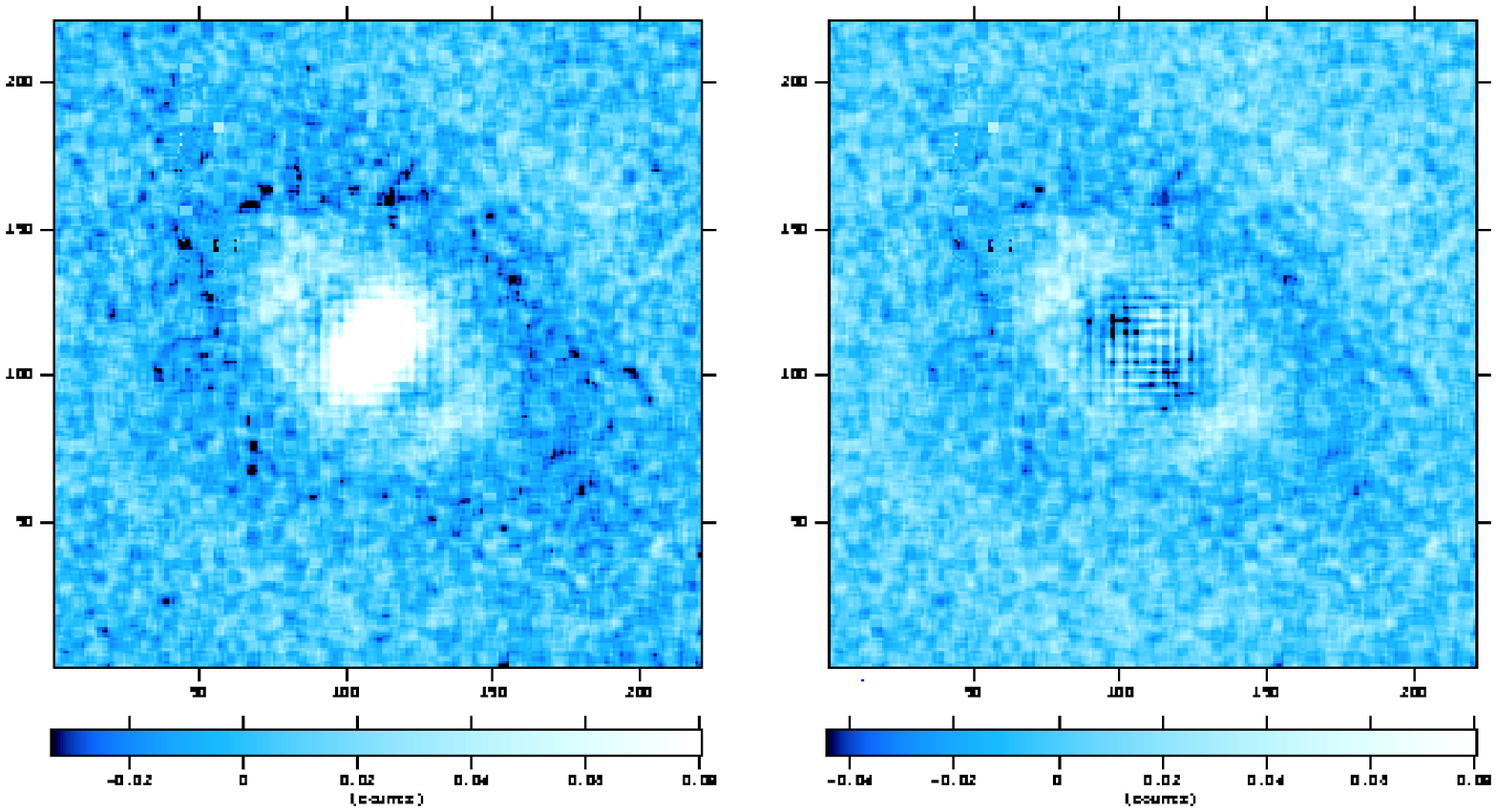}
\caption{Residual images in the \kei\ band (galaxy -- model) of PKS~03120017 (top) and PKS~03120018 
(bottom) obtained applying a model without (on the left) and with (on the right) a central unresolved component.}
\label{residuals}
\end{figure*}

$\bullet$ In two out of the four sources (PKS~03120017 and PKS~03120018), we found clear evidence for an unresolved 
component on top of the elliptical galaxy profile. The \kei band residual images, obtained by 
subtracting the fitted models from the original images, are shown in
Fig. \ref{residuals}; the residuals obtained with 
MODEL~1 are shown in the left panels, while those obtained with MODEL~2 are shown in 
the right panels. 
Introducing an unresolved component improves the quality of the fit, mostly in the \KEI\ band, 
as evident from the residual images and the $\chi^{2}$ values reported in Table~\ref{galfit}. 
The excess with respect to the bulge profile in the innermost region increases towards longer wavelengths. 
The variation in the reduced $\chi^2$, obtained by adding a nuclear component, is significant at a high
($>$ 99.9999\%) confidence level, estimated using an F--test, for both sources.
We note that the $\chi^2$ statistic is computed over a very large 
number of degrees of freedom (see Table \ref{galfit}),  
corresponding to each individual pixel in the image. 
In order to further assess the improvement in the $\chi^2$ statistic 
obtained by the addition of a nuclear component, whose contribution is relevant 
only over a small portion of the image, we ran GALFIT (both MODEL 1 and MODEL 2) 
over the central innermost part of the image (25$\times$25 pixels) keeping fixed 
the best fit parameters obtained by the analysis of the entire image.
Also in this case, the fit quality improves significantly ($>$ 99.9999\% level, according to the F--test) 
using MODEL 2 with respect to MODEL1, indeed enforcing the findings obtained considering a larger region.

\begin{figure*}
\centering
\includegraphics[width=0.5\textwidth]{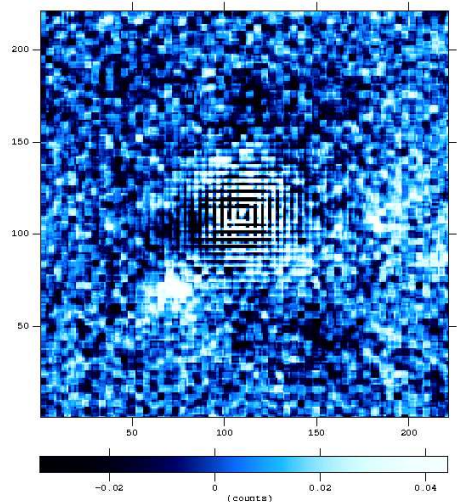}
\caption{Residual image (galaxy -- model) in the \kei\ band of source Abell2690013 obtained applying MODEL~1.}
\label{a269013}
\end{figure*}

$\bullet$ The fitting procedure found no evidence for a unresolved nuclear component in Abell~2690013. The residual 
images obtained with MODEL 1 in both \GEI\ and \KEI\ bands show no need for a central unresolved source in the 
fitting model; the residuals in the \KEI\ band are shown in Fig. \ref{a269013}. 
If a nuclear component is present, its magnitude is fainter than the limit of our observations. The 
absence of a central unresolved component is also supported by the
X-ray analysis: the X-ray emission is well fitted by a thermal model consistent 
with the extended appearance, as shown in Figure \ref{contorni}. 

$\bullet$ Thanks to the high quality of the ISAAC images with sub-arcsec seeing, it has been possible to unveil 
the complex morphology of  source Abell~1835140. It is a close interacting system, composed by two nearby 
bulges embedded in a common envelope 
(see bottom right panel of Fig. ~\ref{xbong_k}). The angular separation of the two 
nuclei is 0.9 arcsec. The morphological fitting was attempted including simultaneously the 
two components, despite the small angular separation. We fitted a S\'ersic model for both galaxies: 
in the \GEI\ band the fit did not provide a statistically reasonable solution (i.e., by means of $\chi^{2}$), 
while in the \KEI\ band we obtained a steep
profile for the northern nucleus and a flatter profile for the southern one. 
Therefore we performed a new fit, adding an unresolved component only to the northern source and a de Vaucouleurs 
profile to both the bulges, as performed for the other \xbongs. The results are suggestive of the presence of a point-like component 
in the northern source of the double system. 
Since the fitting did not provide 
a statistically acceptable description of the two sources, leaving significant residuals because of their low spatial separation, 
this result should be considered just qualitative.

\subsection{Near-Infrared color}

The surface brightness decomposition technique allowed us to separate the galaxy 
from the unresolved nuclear contribution in two out of four \xbongs. 
The host galaxy parameters (m$_{host}$, r$_e$) and the \GEI--\KEI\ colors are in good agreement 
with the \kei band luminosity vs. radius relation (Pahre 1999) and the 
color typical of elliptical galaxies (Cutri et al. 2000; Mannucci et al. 2001), respectively,
lending further support to the morphological analysis. The total luminosity of the galaxies 
in our sample is of the order of 
$L^{\star}$ of early-type galaxies in the \KEI\ band (Kochanek et al. 2001).

\begin{figure}
\centering
\includegraphics[width=0.5\textwidth]{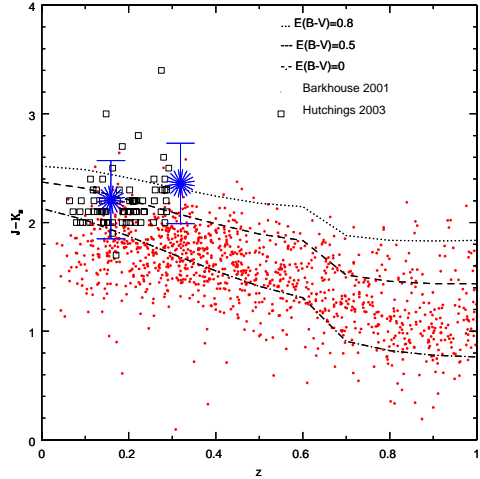}
\caption{Near-infrared color obtained for the AGNs hosted in PKS~03120017 and PKS~03120018 compared with literature 
data. Filled (red) dots are the optically selected quasars with near-infrared counterparts in the 2MASS (Barkhouse et al. 2001), 
while the open (black) squares are the 2MASS selected red quasars from Hutchings et al. (2003). The tracks plotted represent 
the color-redshift relations for a composite quasar template with different extinction values ($E(B-V)$=0, 0.5, 0.8, from bottom to top). }
 \label{color_z}
\end{figure}

The nuclear near-infrared colors of PKS~03120018 (\GEI--\KEI=$2.21\pm0.36$) and PKS~03120017 
(\GEI--\KEI=$2.36\pm0.37$) are reported in
Figure \ref{color_z} along with literature data. Dots represent
the optically selected quasars with near-infrared counterparts in the
2MASS (Barkhouse et al. 2001), while the open squares are the 2MASS
selected red quasars from Hutchings et al. (2003) in a redshift range
(0.1 $< z <$ 0.3) similar to that of our sources.
Even if affected by large uncertainties, mainly due to the extremely faint
\gei band magnitudes, the \GEI--\KEI\ colors of the two \xbongs\ are consistent with
those of 2MASS red quasars at the same redshift. The tracks plotted in Fig. \ref{color_z} represent 
the color-redshift relations for a composite quasar template from the Large Bright Quasar Survey  (LBQS; Francis et al. 1991) extended to the near-IR
using the mean radio-quiet quasar energy distribution (SED) by Elvis et al. 1994 (see also Maiolino et al. 2000). 
We have included the effect of dust
attenuation on the LBQS template using a dust screen model and the
SMC extinction law (Pei at al. 1992), to keep into account the presence of nuclear or galactic dust.  The curves represent different
extinction values ($E(B-V)$=0, 0.5, 0.8 from bottom to top). 
At the face value, the best-fit \GEI--\KEI\ color of the nuclei implies 
$E(B-V) \simeq 0.5 - 0.8$  which corresponds to $N_H \simeq 3 - 5 \times 10^{21}$ cm$^{-2}$ 
if we use a standard Galactic dust-to-gas 
ratio\footnote{$E(B-V)=1.7\times10^{-22}\times N_H$ cm$^{-2}$ (Bohlin et al. 1978).} 
as a conversion factor. %The comparison of the extinction with the gas absorption obtained from the X-ray analysis will be discussed in Section \ref{discussion}.

\section{Simulated optical spectra}
\label{simulspec}
In order to check whether the extinction values, estimated from the nuclear near-infrared colors 
(\GEI--\KEI), could explain the lack of the optical emission lines, we compared the observed 
optical spectra of our \xbongs\ with simulated optical spectra. 
We adopted the SDSS quasar composite spectrum (Vanden Berk et al. 2001) extended to the 
near infrared (see previous section) to reproduce the nuclear component. 

The approach we used is as follows:
\begin{enumerate}
\item using the IRAF task \textit{deredden}, the quasar template was reddened 
with different extinction values: $E(B-V)$=0, 0.2, 0.4, 0.6, 0.8;
\item the templates were redshifted to the redshift of the sources using the IRAF task \textit{dopcor};
\item the spectra were normalized to the \KEI\ band on the basis of the estimated magnitude for the unresolved central source;
\item the nuclear simulated spectra were summed to the observed \xbong\ optical spectra. 
\end{enumerate}
We computed simulated spectra for both PKS~03120017 and PKS~03120018.
In spite of the good signal-to-noise ratio of the observed spectra, the $[OIII]_{5007\AA}$ 
emission line is not visible in all the 
summed spectra, even in the case without reddening. On the contrary, it is possible to reveal 
the presence of an emission line at 
the expected wavelength of the redshifted $H\alpha_{6563\AA}$ in the spectra with $E(B-V)$=0. 

\begin{figure}
 \centering
\includegraphics[width=0.5\textwidth]{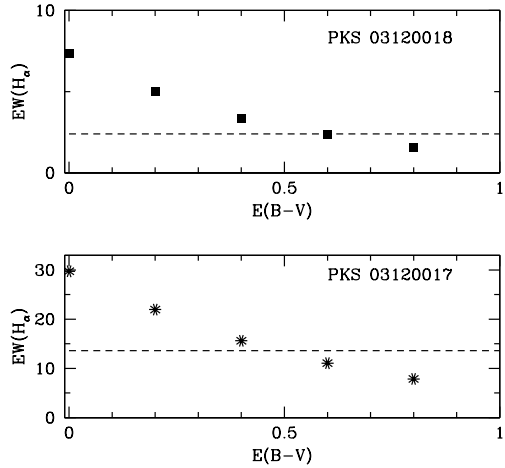}
\caption{The observed equivalent width of the $H \alpha$ line plotted against the $E(B-V)$ of the simulated spectra; filled squares in the upper panel 
represent PKS~03120018 and stars in the lower panel PKS~03120017. The dot-dashed lines represent the EW detection limit curves as estimated following Mignoli et al. (2005).}
 \label{ew}
\end{figure}

In order to quantify the emission-line detection limits as a function of $E(B-V)$, we measured 
the equivalent width (EW) of the $H\alpha$ in the simulated spectra. The average EW of the $H\alpha$ 
line in the SDSS QSO composite spectrum is $\sim$ 190 $\AA$ (Vanden Berk et al. 2001). 
Since the EW is defined as the ratio between the line and the continuum fluxes, its value does not 
change applying the extinction in the nuclear simulated spectra. On the contrary, 
the EW in the total spectra depends from both the nuclear and the steady host galaxy continuum, thus 
it decreases for increasing extinction. 
Figure \ref{ew} shows the equivalent width measured in the summed spectra as a function of the $E(B-V)$  
along with the 3$\sigma$ level of detectability (dot-dashed line) for both sources. 
The EW detection limit has been estimated following the relation presented in Mignoli et al. (2005):  
\begin{equation}
    EW(detection\;limit) = {SL \times \Delta\over{(S/N)_{cont}}}
    \end{equation}  
where $\Delta$ is the resolution element (in \AA) of our spectra, $SL$ is the significance level 
of the detectable line, expressed in terms of $\sigma$ above the continuum noise, and $S/N$ is the 
signal-to-noise ratio of the continuum in the total spectra. 
The extinction values for which the $H\alpha$ equivalent width is lower than the estimated detection 
limit is  $E(B-V) \sim 0.5$, in fair agreement with the values found with the near-infrared morphological 
analysis. In the next section we will compare all the extinction values found so far. 
The above described results are not strongly dependent on the 
assumed $H\alpha$ EW. In fact, also adopting a lower value (70 $\AA$, corresponding to the
2$\sigma$ lower bound of the average intensity of FBQS survey; Brotherton et al. 2001 \footnote{The dispersion around the 
average H$\alpha$ value is not quoted for the SDSS composite spectrum.}), a weak 
line is detected in the simulated spectra with null extinction.

\section{Discussion}
\label{discussion}

The surface brightness decomposition technique allowed us to detect and separate the nuclear from the galaxy contribution in two (PKS~03120018 and PKS~03120017) out of the four \xbongs. 

There is no evidence for the presence of a significant nuclear component in Abell~2690013; the magnitude and the effective radius of the galaxy are in good agreement with the Kormendy relation extended to the \KEI\ band (Pahre 1999). Moreover, the presence of an active nucleus could be ruled out also thanks to a more careful analysis of the X-ray image. The X-ray source is clearly extended (Fig.~\ref{contorni}) and the X-ray spectrum is well fitted by a thermal model.

The presence of two close ($d< $0.9 arcsec) galaxies embedded in a common halo with disturbed morphology in source Abell~1835140 prevented us from a detailed analysis with GALFIT. A qualitative analysis suggested the presence of a unresolved component in the northern source of the system, possibly associated with the X-ray emission. 

The estimated nuclear \GEI--\KEI\ colors, even if affected by large errors due to the uncertainties on 
the nuclear \gei band magnitudes, are consistent with extinction values of $E(B-V) \sim$ 0.5 and 0.8 for PKS~03120018 and PKS~03120017, respectively. For both sources, the lower limit found with the optical simulated spectra, as explained in Section~\ref{simulspec}, is $E(B-V) \sim 0.5$. These values are in agreement with the gas absorption column densities found with the X-ray spectral analysis, assuming a Galactic dust-to-gas ratio (Bohlin et al. 1978). 

\begin{figure}
\centering
\includegraphics[width=0.48\textwidth]{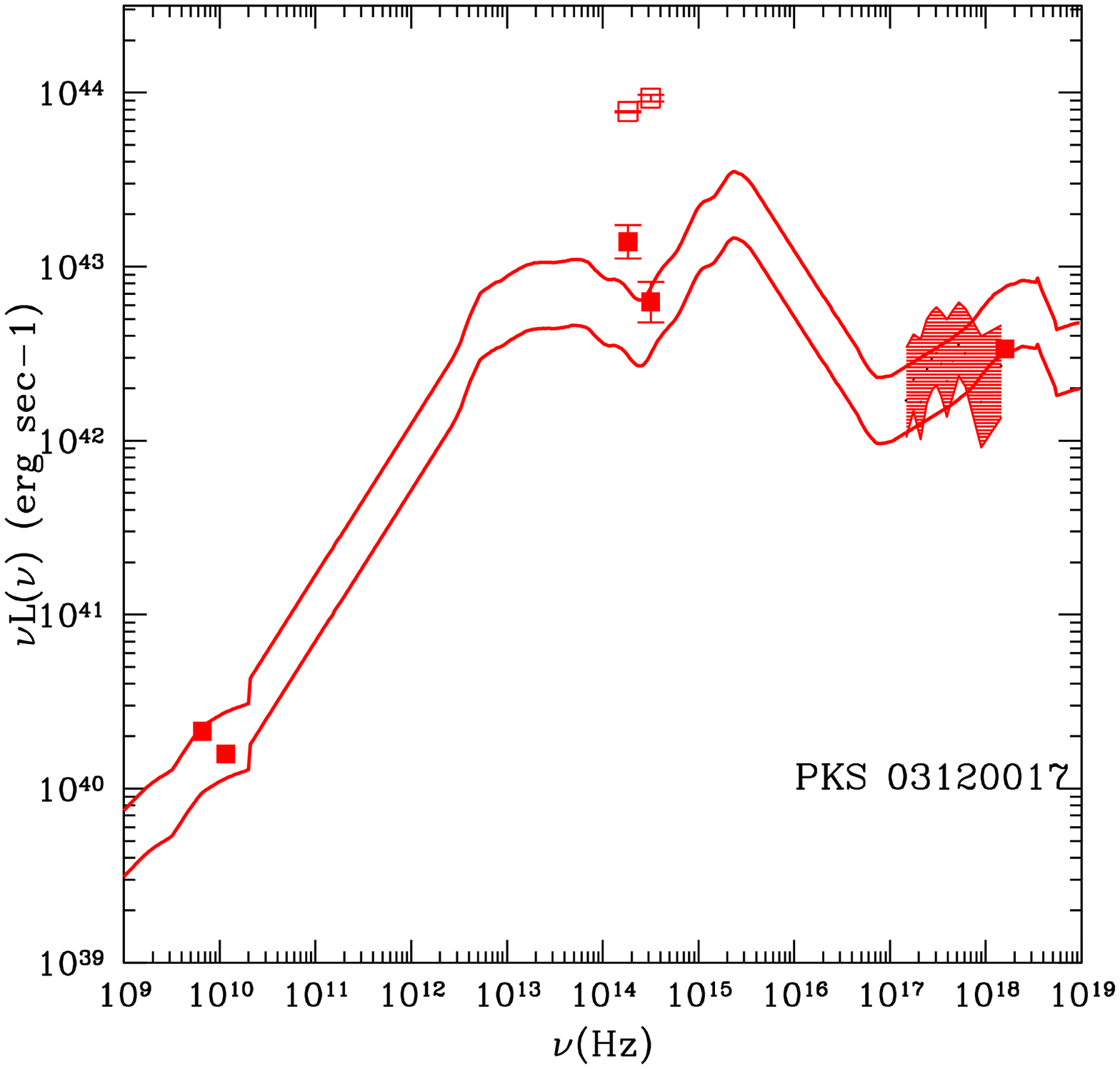}
\includegraphics[width=0.48\textwidth]{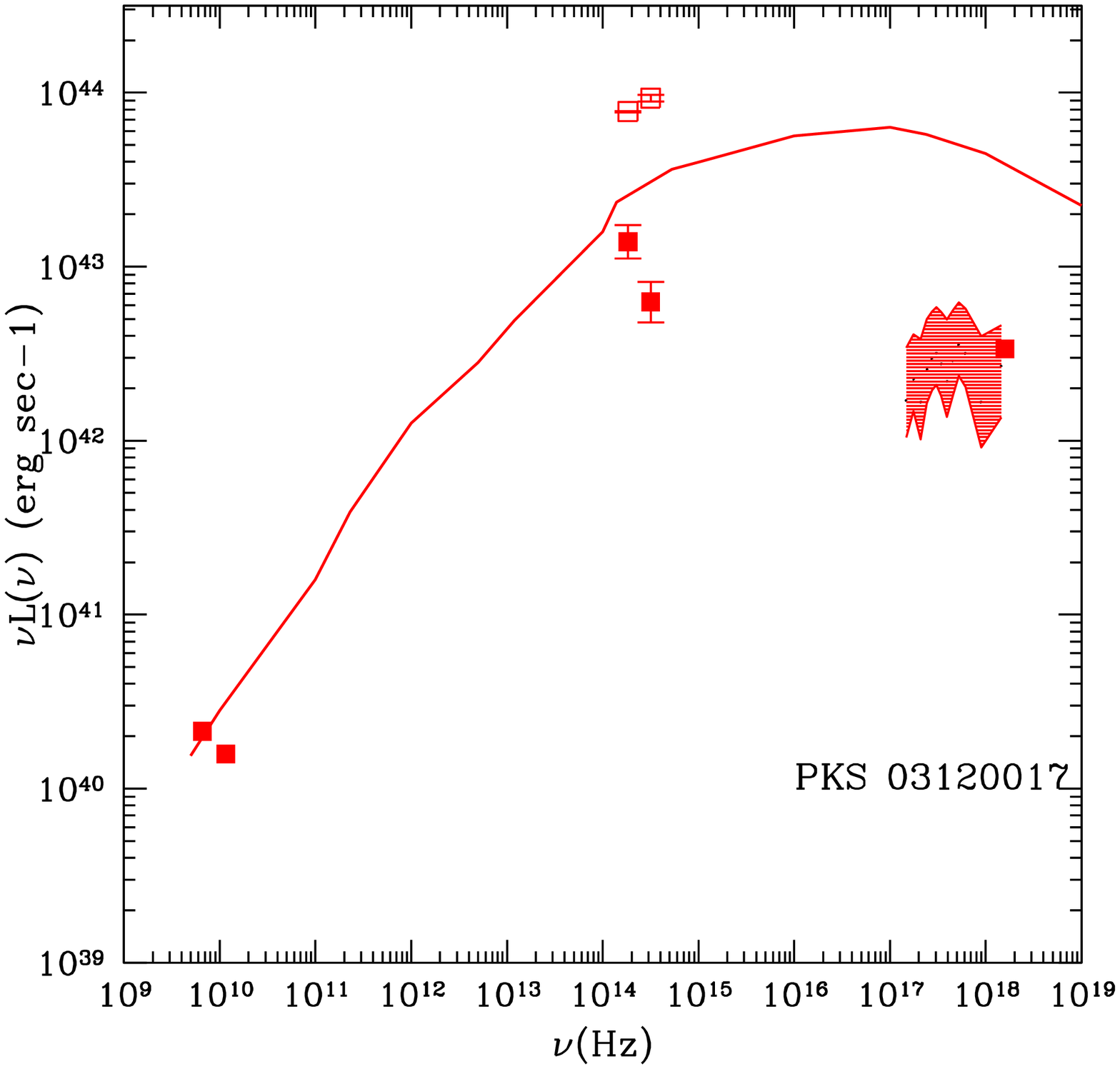}
\caption{The broad-band spectral energy distribution of source PKS~03120017 (filled squares=nuclear data; open squares=host galaxy data) compared with the Type 1 AGN SED (Elvis et al. 1994; left panel) and with a blazar SED with radio luminosity consistent with that of our object (Fossati et al. 1998; right panel). The AGN SED is normalized both to the radio and the X-ray data (rest frame 5 GHz and 5 keV, respectively).}
 \label{sed1}
\end{figure}

\begin{figure}
\centering
\includegraphics[width=0.48\textwidth]{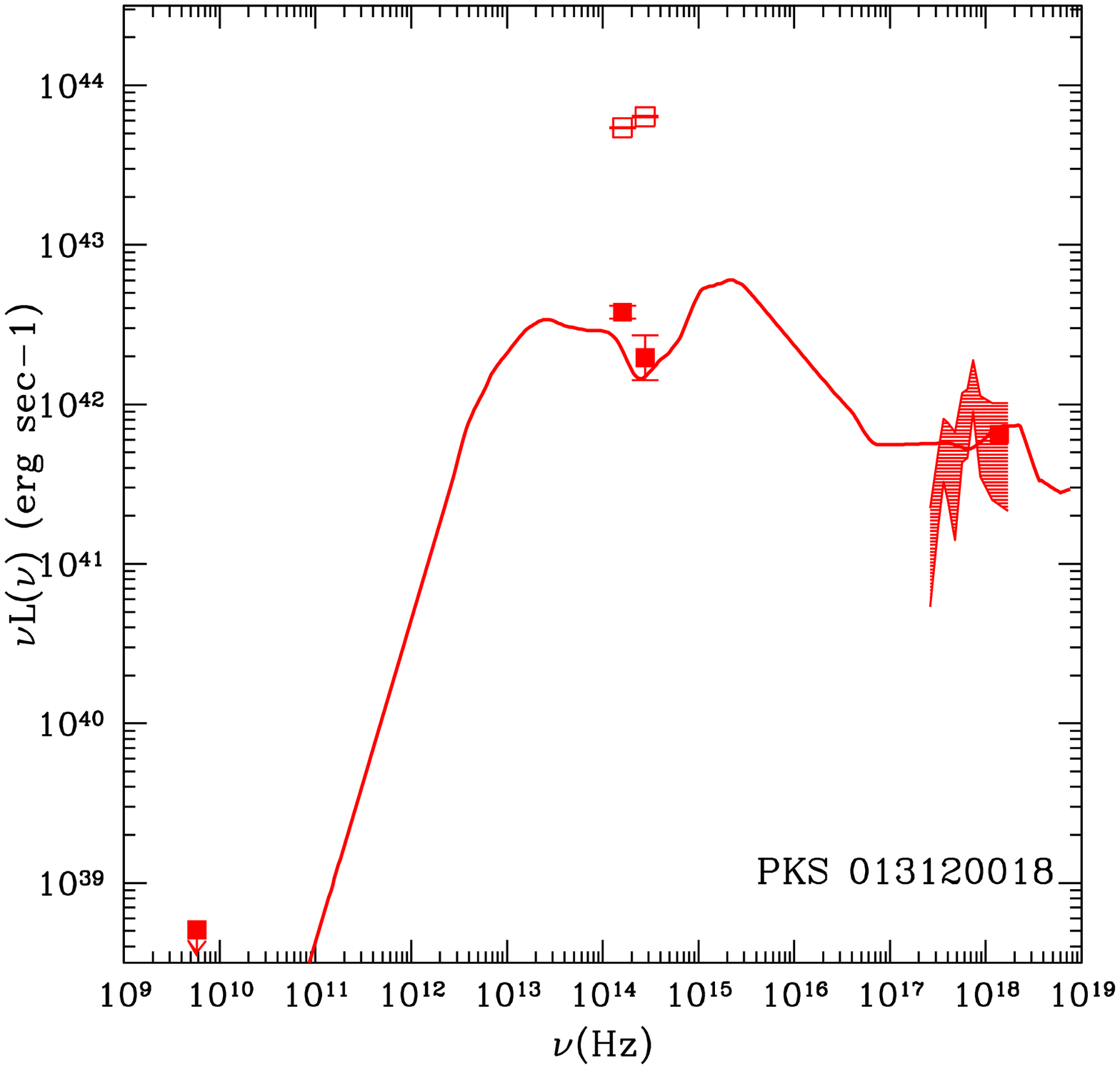}
\includegraphics[width=0.48\textwidth]{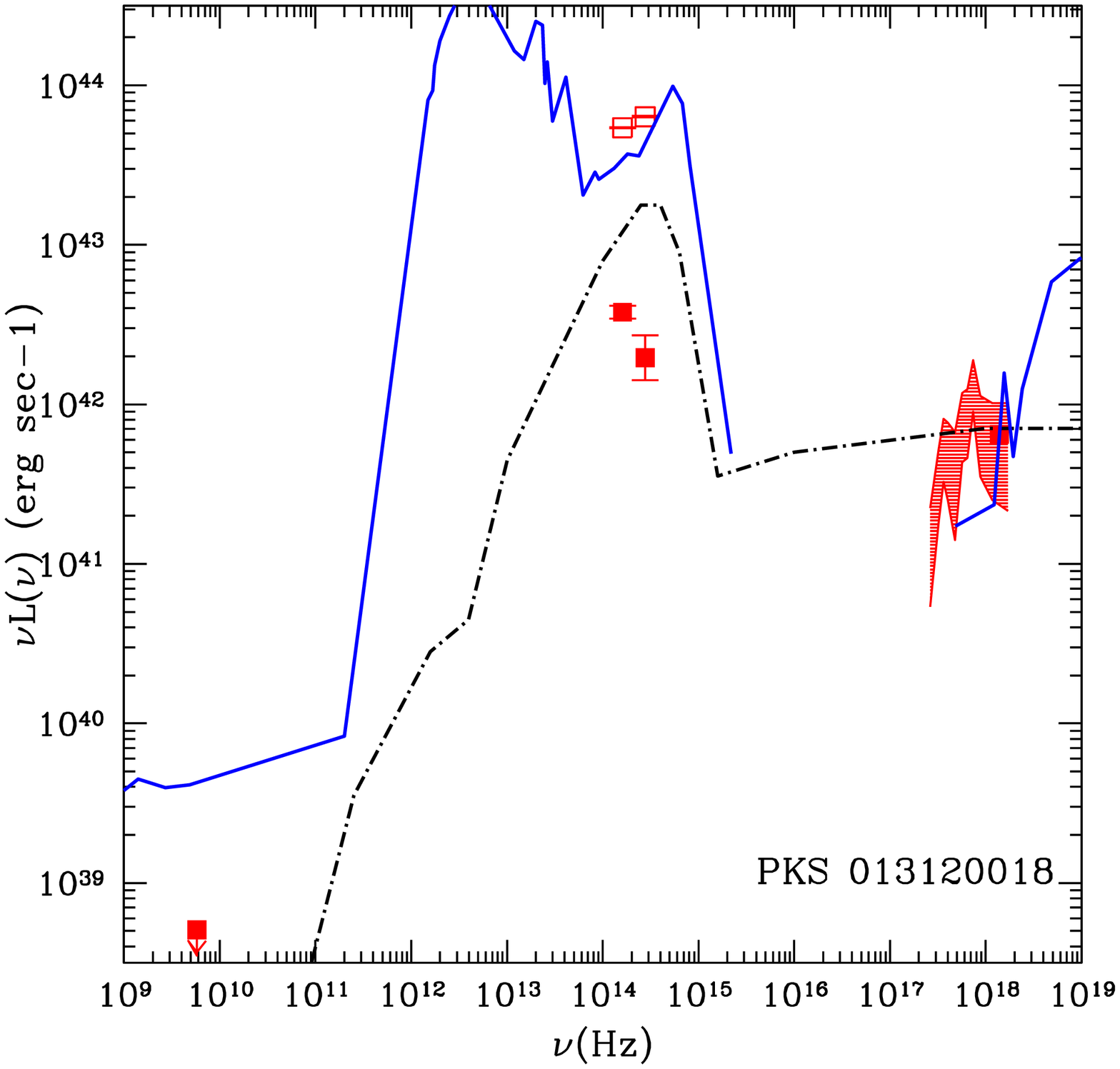}
\caption{The broad-band spectral energy distribution of source PKS~03120018 (filled squares=nuclear data; open squares=host galaxy data) is compared with the Type 1 AGN SED (Elvis et al. 1994) normalized to the X-ray data (5 keV, left panel) and with that of the highly obscured Seyfert 2 galaxy NGC 6240 (continuous line) and with
the RIAF model (dot-dashed line; Yuan and Narayan 2004, right panel).}
 \label{sed2}
\end{figure}

The SEDs of  PKS~03120017 and PKS~03120018 are reported in Figure \ref{sed1} and \ref{sed2}.  Open squares represent the host galaxy and filled squares the nuclear emission. 
The X-ray spectra are the same as Figure \ref{xray-spec}, taking into account the errors on the X-ray flux. We note that source PKS~03120017 has been detected at 5 and 8.7 GHz, with flux densities of 1.29 and 0.55 mJy, respectively.
In the left panels the data are compared with the Type 1 AGN template from Elvis et al. (1994), 
normalized to match the 5~keV rest-frame luminosity for both sources as well as the 5 GHz rest frame luminosity for PKS~03120017. 
The X-ray to near-infrared ratio of both sources is consistent with that of  a Type 1 AGN. 
The radio spectral index of source PKS~03120017 is very steep ($\alpha \sim$ 1.7), at variance with the value expected for an active nucleus ($\alpha \lesssim$ 0.8). However, the radio observations are not simultaneous 
and thus source variability may affect the slope estimated above. 

%Some of the of the alternative possibilities to explain the nature of \xbongs\ presented in the Introduction (Section~\ref{xbongintro})
%Whether the observed broad-band SEDs are consistent with one or more among the alternative possibilities on the nature of \xbongs\ presented in the Introduction, it is discussed in the following.

 The detection of PKS~03120017 in the radio band suggested (Brusa et al. 2003) that it could be an example of a high-energy-peaked BL Lac (the so called HBL). To shed light on the nature of the broad-band emission of this object, we report in Fig.~\ref{sed1} (right panel) the SED with the lowest radio luminosity belonging to the ``blazar sequence" (Fossati et al. 1998), roughly matching the observed 5 GHz flux density of our source. The X-ray-to-radio luminosity ratio of PKS~03120017 is one order of magnitude lower than that expected for a HBL. The HBL SED does not adequately fit the nuclear data of PKS~03120017 suggesting that either the blazar sequence breaks down at low luminosities or, more likely, that PKS~03120017 is not a BL Lac object.

The presence of Compton-thick absorption for source PKS~03120018 was favoured by Comastri et al. (2002), on the basis of the comparison of the source SED (including the host galaxy) with that of NGC~6240, the prototype of this class of objects (see their Fig.~5). The near-infrared nuclear fluxes, obtained with the present analysis, along with 
a re-analysis of the X-ray spectrum, consistent with Compton-thin obscuration (N$_H$$\sim$5 $\times 10^{21}$ cm$^{-2}$), make the Compton-thick hypothesis no more tenable (Fig.~\ref{sed2}, right panel).

A multiwavelength spectral fit to the total emission of source PKS~03120018 with a RIAF plus thin-disk model was presented by Yuan \& Narayan (2004). Even if a detailed analysis of the RIAF model is beyond the scope of this paper, their model is also reported in Fig.~\ref{sed2} (right panel). We note that the solution proposed by Yuan \& Narayan (2004) does not provide a good fit to the nuclear data; however, given the large number of free parameters in RIAF models, we cannot rule out that inefficient accretion may explain the PKS~03120018 SED.  

To explain the properties of more than half of the \xbongs\ selected
in the \chandra\ deep fields at high redshift, Rigby et al. (2006)
proposed the presence of extranuclear dust and gas distributed on
large scales (i.e., kpc dust lanes). Dust lanes have been observed in
HST images of nearby galaxies hosting a Seyfert 2 nucleus (Malkan et
al. 1998) and seem also to be ubiquitous in nearby early-type active
galaxies, selected from the Ho et al. (1995) sample and analyzed by
Lopes et al. (2007). X-ray observations of galaxies belonging to the
Malkan et al. sample suggest a strong correlation between the presence
of dust lanes and X-ray absorption (Guainazzi et al. 2001, 2005).
Moreover, the presence of
extranuclear dust lanes in both samples is always associated with optical emission
lines, at variance with the \xbong\ definition. If
extranuclear dust were responsible of the \xbong\ classification, as
proposed by Rigby et al. (2006), it should be visible in our sources in the
optical band, since our \xbongs\ are much brighter and at lower
redshift. We have retrieved from the HST archive two ACS/HRC images of
source PKS~03120018 taken in the F475W and F625W filters. The optical
images along with near-infrared images do not show any obvious
large-scale dust structure.

If heavy nuclear absorption or absorption related to the presence of
dust lanes are not responsible for the lack of optical emission lines
in PKS~03120017 and PKS~03120018, our results require an alternative
explanation. 
 One speculative possibility is that the ionizing flux
does not reach the narrow line regions because absorbed by intervening
dust and gas, with $E(B-V)=0.5$ and N$_H \sim$10$^{21}$cm$^{-2}$,
covering 4$\pi$ at the central engine.  
%, or simply because the
%central source is weak and not enough UV photons are produced, or a
%combination of both.

Having estimated the host galaxy magnitude, it is possible to compute
the black hole mass adopting the Marconi \& Hunt (2003) relation,
safely assuming that it holds at the redshift of our 
sources. The rest frame \kei band luminosity has been computed using
an evolving elliptical galaxy template \footnote{The galaxy
template includes an exponential star formation with a time-scale
$\tau$=0.3 and formation redshift~$\sim$6.} (Bruzual \& Charlot 2003)
to properly account for the K-correction. The black hole masses
obtained are $6.5 \times 10^8 M_\odot$ and $1.4 \times 10^8 M_\odot$
for PKS~03120017 and PKS~03120018, respectively.  Assuming a range of 
the bolometric correction value (15--30) which, 
following Marconi et al. (2004; see also Elvis et al. 1994), is 
appropriate for the X-ray luminosity of
the \xbong\ nuclei ($5.0 \times 10^{42}$ erg sec$^{-1}$ and $1.1
\times 10^{42}$ erg sec$^{-1}$ for PKS~03120017 and PKS~03120018,
respectively), it is possible to
derive the bolometric luminosity and, therefore, to estimate the
Eddington ratio ($L_{bol} / L_{Edd}$) of the two sources. 
The resulting Eddington ratio is close to $L_{bol} /
L_{Edd} \sim 10^{-3}$ for both sources. 
Should the bolometric correction 
be significantly lower (as it seems to be the case among low luminosity 
AGN e.g. Ho 1999)
then the estimated Eddington ratio would correspondingly
decreases.
At such low values of the Eddington rate, the accretion 
is expected to be radiatively inefficient and described by a
RIAF solution.
However, it has also been suggested that the thin disk solution 
is stable down to Eddington ratios as low as $10^{-6}$ 
(Park \& Ostriker 2001).  Therefore, the lack of optical emission lines
may be explained by a either a RIAF model or by 
standard accretion disk which, at low values of the Eddington ratio,
does not produce enough UV photons able to ionize the
narrow line regions.

\begin{figure}[t] \centering
\includegraphics[width=0.6\textwidth]{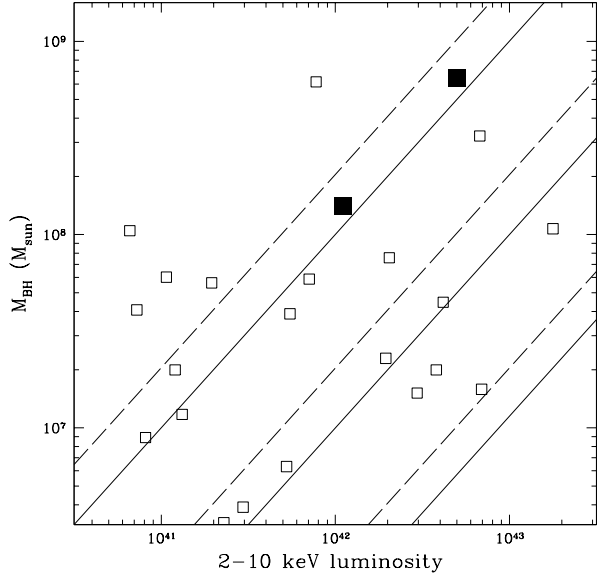}
\caption{Black hole masses (in units of solar mass) versus 2--10 keV
luminosity. Filled squares are the two \xbongs, open squares the LLAGN
from Panessa et al. (2006). The lines represent the expected
correlation between the two plotted quantities for two different
assumptions on the bolometric correction ($k_{bol} = 15$ and $k_{bol}
= 30$, solid and dashed line, respectively) and
L/L$_{Edd}$=0.001--0.1 (from upper left to bottom right).}
 \label{masse}
\end{figure}

In Fig.~\ref{masse} the black hole masses of the two \xbongs\ as a
function of the X-ray luminosity (filled squares) are reported along
with the low-luminosity AGN (LLAGN) from Panessa et
al. (2006). Keeping into account the uncertainties on our
estimates, the black holes that reside in our objects are more massive
than those of the LLAGN of Panessa et al. (2006) and, 
for the same X--ray luminosity, are characterized by a lower Eddington ratio.

\section{Conclusions}

We have presented a multiwavelength analysis of four \xbongs\ selected
from the {\rm HELLAS2XMM} survey. For these sources, deep near-infrared
images taken with ISAAC at VLT, good-quality optical spectra and \xmm\
data are available.  Applying the morphological decomposition
technique, we were able to detect the 
presence of a nuclear component in two out of the four sources (PKS~03120017 and PKS~03120018).
There is no evidence of nuclear emission in the near-infrared in
Abell~2690013; moreover, the X-ray appearance is consistent with an
extended source. For source Abell~1835140, the near-infrared images
reveal a complex morphology, where two sources are embedded in a
common envelope. The main issue about the \xbong\ nature is whether
they represent a truly distinct class or, rather, they are a mixed
source population. Our results point towards the latter hypothesis.

The results regarding the nature of the \xbongs\  with nuclear component and the lack of optical emission lines can be summarized as follows.
\begin{itemize}
\item Source PKS~03120017 and PKS~03120018 are well described by 
a mildly obscured ($E(B-V)= 0.5-0.8$) optically weak nucleus, responsible for the X-ray emission, 
hosted by a bright galaxy (mag$_{K_s}^{nucl}$--mag$_{K_s}^{host}$ $\sim 4$). 

\item The lack of optical emission lines cannot be attributed to observational limitations
such as a inadequate observational setup or low signal-to-noise-ratio, 
at least for the objects in the present sample for which high-quality spectroscopic and photometric 
observations are available.

\item We can safely discard for the two objects both a Compton-thick
scenario (as found by Caccianiga et al. 2007 for a sample of elusive AGN in the 
XBSS) and an important blazar contribution. A RIAF solution 
seems to be supported by the estimated values of the Eddington ratios
in the two objects and is not ruled out by the broad-band SED fitting, 
also because of the large number of free parameters in the RIAF 
model that can be tuned in order to reproduce the observed SEDs.
A weak nuclear source, described by a standard accretion disk solution, but not
powerful enough in the production of UV photons, would also provide 
an acceptable description of the observations.

\item The presence of a thin nuclear gas and dust structure 
(as argued by Cocchia et al. 2007) covering 4$\pi$ at the nuclear source, 
combined with the low level of activity of the BH, could prevent the 
ionization of the narrow-line regions and produce also the extinction we measured.\\

The analysis of already obtained VLT/VIMOS-IFU (Integral Field Unit) observations could 
help in the detection of the typical AGN emission lines down to a very faint flux limit and 
in the study of the gas kinematics.

\end{itemize}

\begin{acknowledgements}
We acknowledge financial contribution from contract 
ASI-INAF I/023/05/0 and PRIN--MIUR grant 2006--02--5203.
The authors thank G. Zamorani, G.C. Perola and P. Ciliegi for useful discussions.
\end{acknowledgements}

\end{document}